\documentclass[11pt]{article}

\newif\ifdebug

\debugtrue

\usepackage[margin=1in]{geometry}

\usepackage[utf8]{inputenc}       
\usepackage{lmodern}               
\usepackage[T1]{fontenc}           
\usepackage[english]{babel}

\usepackage[dvipsnames]{xcolor}
\usepackage{latexsym}             
\usepackage{amsmath}               
\usepackage{amssymb}               
\usepackage{amsfonts}
\usepackage{amsthm}                

\usepackage{amsbsy}      
\allowdisplaybreaks

\usepackage{bbm}

\usepackage{mathtools}
\usepackage{thm-restate}

\usepackage{tikz}                  
\usetikzlibrary{arrows,positioning,shapes}

\usepackage{tikzsymbols}

\usepackage{enumerate}             
\usepackage{enumitem}
\setlist{nolistsep}

\usepackage{xparse}                
\usepackage{hyperref}              
\usepackage{url}

\usepackage{color, xcolor}                 
\definecolor{darkgray}{rgb}{0.15,0.15,0.15} 
\definecolor{lightgray}{rgb}{0.94,0.94,0.94}  
\definecolor{lightlightgray}{rgb}{0.97,0.97,0.97}  
\definecolor{darkred}{rgb}{0.80,0.00,0.00}  
\definecolor{darkgreen}{rgb}{0.00,0.70,0.00}    
\definecolor{darkblue}{rgb}{0.00,0.00,0.70} 

\hypersetup{
  colorlinks   = true, 
  urlcolor     = magenta, 
  linkcolor    = magenta, 
  citecolor    = blue   
}

\usepackage{graphicx}              
\graphicspath{{fig/}}              
\usepackage{caption}               
\usepackage{subcaption}            
\usepackage{multicol}              
\usepackage{float}                 
\usepackage{framed}                
\usepackage[framemethod=tikz]{mdframed}

\usepackage{enumitem}              
\usepackage{glossaries}

\usepackage[title]{appendix}

\usepackage[bottom]{footmisc}

\numberwithin{equation}{section}

\theoremstyle{plain}  
\newtheorem{thm}{Theorem}[section]
\newtheorem*{thm*}{Theorem}
\newtheorem{lem}[thm]{Lemma}
\newtheorem*{lem*}{Lemma}

\newtheorem*{claim*}{Claim}
\newtheorem{prop}[thm]{Proposition}
\newtheorem*{prop*}{Proposition}
\newtheorem{cor}[thm]{Corollary}
\newtheorem*{cor*}{Corollary}

\theoremstyle{plain}

\theoremstyle{definition}
\newtheorem{defn}[thm]{Definition}
\newtheorem*{defn*}{Definition}

\newtheorem*{notation*}{Notation}

\newtheorem*{problem*}{Problem}

\newtheorem*{question*}{Question}

\theoremstyle{remark}  
\newtheorem{rmk}[thm]{Remark}
\newtheorem*{rmk*}{Remark}
\newtheorem{ex}[thm]{Example}
\newtheorem*{ex*}{Example}

\usepackage{lastpage,fancyhdr}
\providecommand*{\hr}[1][class-arg]{
    \hspace*{\fill}\hrulefill\hspace*{\fill}
    \vskip 0.65\baselineskip
}

\newcommand{\norm}[1]{\left\lVert#1\right\rVert}

\newcommand{\abs}[1]{\left| #1 \right|}

\NewDocumentCommand{\prob}{mg}{
    \IfValueTF{#2}{
        P(#1 \,\vert\, #2)
    }{
        P(#1)
    }
}

\newcommand\restr[2]{{
  \left.\kern-\nulldelimiterspace 
  #1 
  \right|_{#2} 
  }}

\let\latexchi\chi
\makeatletter
\renewcommand\chi{\@ifnextchar_\sub@chi\latexchi}
\newcommand{\sub@chi}[2]{
  \@ifnextchar^{\subsup@chi{#2}}{\latexchi^{}_{#2}}
}
\newcommand{\subsup@chi}[3]{
  \latexchi_{#1}^{#3}
}
\makeatother

\newcommand{\eps}{\varepsilon}

\newcommand{\N}{\mathbb{N}}

\newcommand{\C}{\mathbb{C}}

\newcommand{\bra}[1]{\langle#1|}
\newcommand{\ket}[1]{|#1\rangle}

\makeatletter
\newcommand\RedeclareMathOperator{
  \@ifstar{\def\rmo@s{m}\rmo@redeclare}{\def\rmo@s{o}\rmo@redeclare}
}

\newcommand\rmo@redeclare[2]{
  \begingroup \escapechar\m@ne\xdef\@gtempa{{\string#1}}\endgroup
  \expandafter\@ifundefined\@gtempa
     {\@latex@error{\noexpand#1undefined}\@ehc}
     \relax
  \expandafter\rmo@declmathop\rmo@s{#1}{#2}}

\newcommand\rmo@declmathop[3]{
  \DeclareRobustCommand{#2}{\qopname\newmcodes@#1{#3}}
}
\@onlypreamble\RedeclareMathOperator
\makeatother

\RedeclareMathOperator{\Re}{Re}

\DeclareMathOperator*{\E}{\mathbb{E}}

\usepackage{braket}

\newcommand{\cG}{\mathcal{G}}

\newcommand{\bM}{\mathbb{M}}

\newcommand{\cS}{\mathcal{S}}

\newcommand{\cV}{\mathcal{V}}

\newcommand{\bigO}{\mathcal{O}}

\usepackage[affil-it]{authblk}

\newcommand{\rhomh}{\rho^{-1/2}}
\newcommand{\rhomo}{\rho^{-1}}

\begin{document}
\title{Gap-preserving reductions and RE-completeness of\\ independent set games}

\author[1]{Laura Mančinska\thanks{mancinska@math.ku.dk}}
\author[1]{Pieter Spaas\thanks{pisp@math.ku.dk}}
\author[1]{Taro Spirig\thanks{tasp@math.ku.dk}}
\author[2]{Matthijs Vernooij \thanks{M.N.A.Vernooij@tudelft.nl}}
\date{}
\affil[1]{Department of Mathematical Sciences, University of Copenhagen}
\affil[2]{Delft Institute of Applied Mathematics, TU Delft}

\maketitle

\vspace{-1cm}
\begin{abstract}
    In complexity theory, gap-preserving reductions play a crucial role in studying hardness of approximation and in analyzing the relative complexity of multiprover interactive proof systems. In the quantum setting, multiprover interactive proof systems with entangled provers correspond to gapped promise problems for nonlocal games, and the recent result MIP$^*$=RE \cite{ji2020mipre} shows that these are in general undecidable. However, the relative complexity of problems within MIP$^*$ is still not well-understood, as establishing gap-preserving reductions in the quantum setting presents new challenges.

    In this paper, we introduce a framework to study such reductions and use it to establish MIP$^*$-completeness of the gapped promise problem for the natural class of independent set games. In such a game, the goal is to determine whether a given graph contains an independent set of a specified size. We construct families of independent set games with constant question size for which the gapped promise problem is undecidable. In contrast, the same problem is decidable in polynomial time in the classical setting. 
    
    To carry out our reduction, we establish a new stability theorem, which could be of independent interest, allowing us to perturb families of almost PVMs to genuine PVMs.
\end{abstract}

\section{Introduction}

Interactive proof systems are a central concept in computational complexity theory. Shortly after the introduction of the multi-prover variant, Babai, Fortnow, and Lund characterized its power by showing that MIP = NEXP \cite{BFL91}, where MIP is the class of languages that can be decided by a multi-prover interactive proof system and NEXP is the set of languages with exponential-time checkable proofs. In fact, already MIP(2,1) = NEXP i.e.\ only two provers and a single round of interaction suffice to reach NEXP \cite{FL92}. 
Such single-round proof systems can be viewed as families of nonlocal games $C=\{\cG_x\}_x$, where the all-powerful provers attempt to convince the efficient verifier that the word $x\in \{0,1\}^*$ belongs to a language $L$, see Section \ref{ssec:games} for a definition of nonlocal games. 
The power of MIP(2,1) is captured by the \emph{gapped promise problem} for the family of games $C$: for fixed constants $0<s< c\leq 1$, given a game $\cG_x\in C$, decide whether the value of $\cG_x$ is at least $c$, or less than $s$, given the promise that one of the two must hold. We denote this as the $(c,s)$-gap problem for $C$ and call $c$ and $s$ the completeness and soundness parameter respectively. 
Importantly, the families of games considered in the above results are \emph{succinctly presented}, i.e., while the question and answer sets can grow exponentially in $\abs{x}$, there are efficient Turing machines $S$ and $V$ to respectively sample questions and verify the answers from the provers, see Section \ref{ssec:games} for a formal definition. 
In fact, the PCP theorem -- the corner stone for the classical theory of hardness of approximation -- can be viewed as a scaled-down version of MIP = NEXP, where the families of games have question and answer sets which grow at most polynomially in $\abs{x}$ \cite{ALM+98,AS98}.

Since the provers are assumed to be all-powerful, it is natural to allow them to take advantage of quantum physics and most notably shared entanglement. The complexity class MIP$^*$, where the provers can share entanglement, is captured by the gapped promise problem defined above where we replace the value of a game by its quantum value \cite{CHTW04}. We will refer to this as the $(c,s)$-gap$^*$ problem. In comparison to MIP, its quantum counterpart MIP$^*$ proved much harder to understand. Over the years, MIP$^*$ was shown to contain MIP and increasingly larger complexity classes \cite{ItoVidick12,Vid16,Vid16err,Ji16,Ji17,NV18a,NV18b,FJVY19,NatarajanWright19}. These developments were further re-enforced by a result of Slofstra who showed that the gapless problem $(1,1)$-gap$^*$ is in general undecidable \cite{Slo19}. Subsequently, Ji, Natarajan, Vidick, Wright and Yuen announced the major breakthrough result that MIP$^*$ = RE, where RE is the class of recursively enumerable languages \cite{ji2020mipre}. More precisely, for any $0<s<1$ they construct a family of so-called \emph{synchronous} games for which the $(1,s)$-gap$^*$ problem decides the Halting problem.\footnote{Note that for any $0<s<c\leq 1$, RE-hardness of the $(1,s)$-gap$^*$ problem for a family of games immediately implies RE-hardness of the $(c,s)$-gap$^*$ problem for the same family of games. We also remark that it was shown in \cite{MNY20} that the complexity of the gapless problem is complete for the class $\Pi_2^0$, a class at the second level of the arithmetical hierarchy which contains both RE and coRE, i.e.\ there is a family of games for which the associated $(1,1)$-gap$^*$ problem is $\Pi_2^0$-complete.} This result indicates that, in general, approximating the quantum value of synchronous games within any additive constant is undecidable. However, this undecidability result does not imply that the quantum value is uncomputable for more specific classes of games. In fact, there are known classes of games for which the quantum value can be computed efficiently \cite{Tsirelson87,Tsirelson93}. It is important to further explore which classes of nonlocal games allow for efficient computation or approximation of the quantum value. Indeed, understanding this could help identify scenarios where quantum entanglement offers an advantage over classical technologies and could lead to the development of new protocols for distributed tasks, such as delegating quantum computation \cite{Gril19,KLVY23}.

The general undecidability result of MIP$^*$=RE has also had significant mathematical consequences, most notably the resolution in the negative of the Connes embedding problem. Showing that it remains RE-hard to compute or approximate the quantum value of more specific classes of games could resolve other long-standing open problems as well, such as proving the existence of a non-hyperlinear group \cite{PaddockSlofstra23}. In fact, a similar undecidability result for so-called tailored non-local games recently led to the resolution of the Aldous--Lyons conjecture \cite{Bowen24a,bowen24b}. 

In short, the complexity of approximating the quantum value of general nonlocal games has now been established. It is therefore the right time to embark on a research program to understand the following general yet more fine-grained questions, which motivate our work:

\begin{center}\emph{What is the complexity landscape of problems in $\mathrm{MIP}^*$? What is the complexity of the $(c,s)\textit{-}\mathrm{gap}^*$ problem for different natural classes of nonlocal games?}
\end{center}

\noindent To answer the above fundamental questions, we must understand the following key points:
\begin{enumerate}[label=\Roman*.]
    \item\label{introkeypoint1} What are natural well-structured classes of games which are MIP$^*$/RE-complete?\footnote{This is analogous to NP-completeness in the classical setting. Recall that, say, 3SAT being NP-complete is a shorthand for: deciding satisfiability of 3SAT instances is NP-complete. Similarly a class of games is MIP$^*$-complete means: for some $0<s<c\leq 1$ deciding the $(c,s)$-gap$^*$ problem for that class of games is MIP$^*$-complete. By MIP$^*$=RE \cite{ji2020mipre}, this is equivalent to RE-completeness.} 

    \item\label{introkeypoint2} How does the complexity of the $(c,s)$-gap$^*$ problem depend on the class of games and the choice of the completeness and soundness parameters  $c$ and $s$? In particular, is there a sharp transition from easy to undecidable in terms of the choice of $c,s$ for certain classes of games?

    \item\label{introkeypoint3} How does limiting the question and/or answer size of a chosen family of games affect the complexity of the corresponding gapped promise problem?
\end{enumerate}

\noindent There are two main approaches to addressing these questions: an algorithmic approach and a complexity-theoretic one. Our results focus on the latter. Before presenting them, we briefly review the current state of the art.
\,\\

\noindent \textbf{Algorithms. } The first approach is to devise efficient algorithms to compute or approximate the quantum value of certain games, which is particularly relevant for question~\ref{introkeypoint2} above. There are only a few examples of classes of games for which we have efficient algorithms to compute or approximate the quantum value. By a result due to Tsirelson \cite{Tsirelson87,Tsirelson93}, the quantum value of so-called XOR games can be computed efficiently by a semi-definite program (SDP). This work subsequently inspired the development of SDP-based approximation algorithms for several more general classes of games, including unique games and more recently 3-Coloring \cite{KRT10,beigi2011lower,CMS24}. So far, none of the existing approximation algorithms in the quantum setting have been shown to be optimal, in the sense that the existence of a better approximation algorithm would imply the collapse of certain complexity classes. In contrast, many classical approximation algorithms are known to be optimal unless P=NP. Showing quantum analogs of such results requires a better understanding of \emph{hardness reductions} in the quantum setting, which we discuss next.
\,\\

\noindent \textbf{Complexity theory (hardness). } The second approach is to study the hardness of gapped promise problems and to relate them to each other, which is relevant for all three questions \ref{introkeypoint1}--\ref{introkeypoint3}.

By \cite{ji2020mipre}, synchronous games are MIP$^*$-complete in a strong sense: the $(c,s)$-gap$^*$ for the class of synchronous games is MIP$^*$-complete \emph{for any} $0<s<c\leq 1$. This result has since seen several refinements. Firstly, as mentioned above, it is now known that the
same complexity theoretic result holds for a subclass of synchronous games called tailored games \cite{bowen24b}. Secondly, synchronous games with either constant question size \cite{NZ23} or constant answer size \cite{Dong24} remain MIP$^*$-complete in a weaker sense\footnote{These classes of games are MIP$^*$-complete in the following sense: for every choice of $0<s<c\leq 1$, there is a family of games with either constant question size or constant answer size such that the $(c,s)$-gap$^*$ problem associated to it is MIP$^*$-complete. This does not mean that the same family of games (with fixed question size or fixed answer size) is MIP$^*$-complete in the strong sense. In fact, it follows from a simple argument that this cannot be the case, see the discussion in Remark \ref{rmk:trivialsoundness}.}, providing a partial answer to question \ref{introkeypoint3} above. Reducing the size of both the question and answer sets to at most a polynomial in the instance size is a major open question in the field, and is closely related to the quantum games PCP conjecture \cite{natarajan24}.
Both of the results \cite{NZ23} and \cite{Dong24} are proven by reducing the Halting problem to a gapped promise problem for the class of games considered. To identify further MIP$^*$-complete classes of games, it is natural to consider reductions \emph{between} gapped promise problems, instead of reductions from the Halting problem directly. This parallels the classical literature on hardness of approximation: for instance hardness of approximating 3SAT is proven by a reduction from another ``stronger'' hardness of approximation result, namely the PCP theorem \cite{hastad}.

More precisely, a reduction from the $(c,s)$-gap$^*$ problem for a family of games $C$ to the $(\overline{c},\overline{s})$-gap$^*$ problem for a family of games $\overline{C}$ is a mapping $C\ni\cG\mapsto \overline{\cG}\in \overline{C}$ satisfying
\begin{itemize}
    \item (completeness) if $\omega^*(\cG)\geq c$, then $\omega^*(\overline{\cG})\geq \overline{c}$, and
    \item (soundness) if $\omega^*(\cG)<s$, then $\omega^*(\overline{\cG})<\overline{s}$;
\end{itemize}
moreover this mapping should be efficient in an appropriate sense, see Section \ref{ssec:complexity} for a formal definition. Crucially, such a reduction implies that if the $(c,s)$-gap$^*$ problem for $C$ is hard, then the $(\overline{c},\overline{s})$-gap$^*$ problem for $\overline{C}$ is also hard. 
To establish a reduction in the quantum setting, one usually needs to construct a quantum strategy for $\overline{\cG}$ given a quantum strategy for $\cG$ and vice-versa. This is greatly simplified when the given quantum strategy for $\cG$, respectively $\overline{\cG}$, is perfect. In this case, the operators from the given strategy will satisfy certain algebraic relations exactly (e.g. orthogonality or commutation relations) which can often be leveraged to directly construct a quantum strategy for $\overline{\cG}$, respectively $\cG$. 
Thanks to this simplification, several \emph{gapless} reductions are known, meaning reductions between $(1,1)$-gap$^*$ problems for various classes of games \cite{ji2013binary,MRV15,Harris24a,Atserias}. For instance, Ji showed that the $(1,1)$-gap$^*$ problem for 3SAT games (in the clause-variable model) reduces to the $(1,1)$-gap$^*$ problem for 3-Coloring games \cite{ji2013binary}, and Man\v{c}inska, Roberson, and Varvitsiotis showed that the $(1,1)$-gap$^*$ problem for synchronous games reduces to the $(1,1)$-gap$^*$ problem for \emph{independent set} games, which we define shortly. 
On the other hand, when the starting strategy is not perfect, its operators will usually only satisfy certain algebraic relations approximately (e.g., they are only approximately orthogonal). In order to construct a quantum strategy for the target game, it is then often required to establish so-called \emph{stability theorems}. Such theorems assert that if a set of operators approximately satisfies certain algebraic relations, then there exists a close-by set of operators that satisfies those relations exactly. For our application, the notion of closeness in a stability theorem must be well-behaved. More specifically, it should scale favorably with the number of operators involved -- a property that is often highly non-trivial to establish. We refer to the discussion of our proof techniques below for further details. Such \emph{quantitative} stability results usually lie at the heart of \emph{gap-preserving} reductions in the quantum setting, i.e., reductions from a $(c,s)$-gap$^*$ problem to a $(\overline{c},\overline{s})$-gap$^*$ problem with $s<c$ and $\overline{s}<\overline{c}$. 

Our main result shows how to lift the gapless reduction of Man\v{c}inska, Roberson, and Varvitsiotis to a gap-preserving reduction. To do so, we prove a new stability theorem, which we believe to be of independent interest, and which could be used to prove further gap-preserving reductions and elucidate the three questions \ref{introkeypoint1}--\ref{introkeypoint3} discussed above.
\,\\

\noindent \textbf{Our results. } The family of games we reduce to consists of independent set games. Given a graph $X$ and a positive integer $t$, the provers in the $t$-independent set game for $X$ aim to convince the verifier that the graph $X$ has an independent set of size $t$, i.e., a set of $t$ pairwise non-adjacent vertices. The question set $[t]=\{1, 2, \ldots, t\}$ consists of $t$ labels, and the answer set $\mathcal{V}(X)$ consists of the vertices of $X$. To play the game, the verifier samples a pair $(i,j)$ from $[t]\times [t]$ according to some probability distribution, e.g., uniformly at random (see Definition~\ref{def:indsetgame} for the other probability distribution we will be using), and sends $i$ to the first prover and $j$ to the second. The provers then respond with vertices $u,v\in \mathcal{V}(X)$ and they win if and only if the following is satisfied:
\begin{itemize}
    \item if $i=j$, then $u=v$, and
    \item if $i\neq j$, then $u$ and $v$ are distinct, non-adjacent vertices.
\end{itemize}
Our main technical result is the following.

\begin{restatable}{main}{mainreduction}
\label{thm:mainreduction}
    Let $\cG$ be a synchronous game with uniform question distribution. Then there is a $t$-independent set game $\overline{\cG}$, where $t$ is the number of questions of $\cG$, such that the following hold:
    \begin{itemize}
        \item if $\omega^*(\cG)=1$, then $\omega^*(\overline{\cG})=1$, and
        \item if $\omega^*(\cG)< 1-\eps$, then $\omega^*(\overline{\cG})< 1-\bigO(\frac{\varepsilon^8}{t^{4}})$.
    \end{itemize}
\end{restatable}

As stated, this theorem does not automatically give a gap-preserving reduction from any given class of synchronous games $C=\{\cG_x\}_x$ to the class of independent set games. Indeed, the gap $\bigO(\frac{\varepsilon^8}{t_x^{4}})$ depends on the number of questions $t_x$ of the games $\cG_x$ which could increase with $\abs{x}$. However, as alluded to above, Natarajan and Zhang constructed in \cite{NZ23} an RE-complete family of synchronous games with question sets of constant size, uniform question distribution, and with answer sets growing quasi-polynomially in $\abs{x}$. Using this result, the following corollary then follows easily from Theorem~\ref{thm:mainreduction} (see Section~\ref{sec:analysis_approx}):

\begin{restatable}{mcor}{introhardness}
\label{cor:intro_hardness}
There exist a constant $0<s<1$, a positive integer $t$, and a family $C$ of $t$-independent set games such that the $(1,s)$-gap$^*$ problem for $C$ is $\mathrm{RE}$-complete.
\end{restatable}

Our results have several implications. First, they address the key points \ref{introkeypoint1}--\ref{introkeypoint3} outlined above as follows:
\begin{enumerate}[label=\Roman*.]
    \item We show that independent set games are MIP$^*$/RE-complete. This class of games is a natural well-structured class of games which also plays a significant role in (classical) complexity theory. 
    \item We demonstrate that there exists a constant $0<s<1$ for which the $(1,s)$-gap$^*$ problem for independent set games is RE-hard and another constant for which it becomes easy, in fact trivial, see Remark \ref{rmk:trivialsoundness}. However, the precise constant (if any) at which the problem transitions from RE-hard to easy remains unknown, see the future directions below for further details.
    \item Our MIP$^*$-complete class of games has constant question size, and is a more structured class than the class of synchronous games with constant question size from \cite{NZ23}. This added structure opens the possibility of applying classically known techniques to compress the graphs of our independent set games, offering a new avenue for tackling the quantum games PCP conjecture.
\end{enumerate}

\noindent Another intriguing implication arises from comparing our problem to its classical counterpart. The family of independent set games in Corollary~\ref{cor:intro_hardness} is succinctly presented, and the vertex sets (i.e., answer sets of the games) grow quasi-polynomially in $\abs{x}$. By an exponential-time reduction, it follows that for the soundness parameter $0<s<1$ from Corollary \ref{cor:intro_hardness}, the $(1,s)$-gap$^*$ problem is undecidable for (non-succinctly presented) $t$-independent set games, whose vertex sets grow polynomially in $\abs{x}$.\footnote{By a similar exponential-time reduction NP equals NEXP. Since we are only asking whether the problem is decidable or not, we can consider such reductions, see also \cite{CulfMastel24}.} In contrast, the same problem is decidable in polynomial time in the classical setting. Indeed, given a graph with vertex set $\cV$, it suffices to check if there is an independent set among the $t$-sized subsets of $\cV$, and there are only $\binom{\abs{\mathcal{V}}}{t}$ such subsets. 

This is a striking phenomenon which adds to the list of quantum complexity-theoretic problems behaving entirely differently from their classical counterparts. Other examples include XOR games, which are NP-hard classically \cite{hastad}, but as mentioned above, become easy in the quantum setting, and linear system games, for which the $(1,1)$-gap problem is efficiently decidable, but the $(1,1)$-gap$^*$ problem is undecidable \cite{Slo19}. Our results show an even more surprising phenomenon: let $t$ be as in Corollary \ref{cor:intro_hardness}. Then for $t$-independent set games, the $(1,1)$-gap problem is efficiently decidable but the $(1,s)$-gap$^*$ problem is undecidable for some constant $0<s<1$. This forms an additional motivation to study questions \ref{introkeypoint1}--\ref{introkeypoint3}, as it gives us a finer insight into the counterintuitive power of entanglement.

Finally, we remark that our proofs also apply to the \emph{commuting operator} value instead of the quantum value, see Section \ref{ssec:games} for a definition. Hence, if the conjecture that MIP$^{co}$=coRE holds, our results would additionally imply that the same gapped promise problem as in Corollary~\ref{cor:intro_hardness} for the commuting operator value would be coRE-complete.
\,\\

\noindent \textbf{Proof techniques.} Next, we briefly outline the main ingredients and the structure of the proof of our main Theorem \ref{thm:mainreduction}. Let $\cG$ be a synchronous game as in the statement of the theorem. Let $Q$ denote its question set, $A$ its answer set, and $V$ its predicate, see Section \ref{ssec:games} for definitions. The \textit{game graph} $X(\cG)$ associated to $\cG$ is defined as follows (see Definition \ref{def:gamegraph} and \cite{MRV15}): The vertices of $X(\cG)$ correspond to pairs $(q,a)\in Q\times A$, and two vertices $(q,a)$ and $(q',a')$ are adjacent if and only if $V(q,q';a,a')=0$ or $V(q',q;a',a)=0$. We then let $\overline{\cG}$ in Theorem \ref{thm:mainreduction} be the $\abs{Q}$-independent set game for the graph $X(\cG)$, with its diagonally weighted question distribution, see Definition~\ref{def:indsetgame}.

The proof now amounts to showing that the two bullet points in the statement of Theorem~\ref{thm:mainreduction} are satisfied. The first bullet point is essentially taken care of by the gapless reduction of \cite{MRV15}, see Section~\ref{sec:analysis_perfect}.

The second bullet point is more involved, and we prove it in the contrapositive: Assuming we have a quantum strategy $\overline{\cS}$ for the $\abs{Q}$-independent set game for the graph $X(\cG)$ which wins with probability greater than $1-\bigO(\frac{\varepsilon^8}{t^{4}})$, we construct a strategy $\cS$ which wins the synchronous game $\cG$ with probability greater than $1-\eps$. To construct the strategy $\cS$, we need to take sums of operators from the strategy $\overline{\cS}$. If the strategy $\overline{\cS}$ were perfect, these sums of operators would directly form a valid quantum strategy. In particular, they would form a Projective Valued Measurement (PVM), i.e., a set of projections summing to the identity. However, as the strategy $\overline{\cS}$ is only close to perfect, the resulting operators are only approximate projections which approximately sum to the identity. We thus need to round these sums of operators to actual PVMs, which we call \emph{rounded} operators, in order to complete the proof.\footnote{In fact, for technical reasons, we actually round the sets of operators which we sum over and not the sums of operators. We refer to Section~\ref{sec:analysis_approx} for more details.} This is where the following stability theorem comes into play, where for a tracial von Neumann algebra $(M,\tau)$, we define the two-norm $\norm{x}_2^2\coloneqq \tau(x^*x)$ for $x\in M$; see Section~\ref{ssec:vNa} for more background on von Neumann algebras. If one is mainly interested in the quantum value, i.e., finite-dimensional strategies, then one can safely assume that $M$ is a matrix algebra with its usual (normalized) trace.

\begin{restatable}{main}{introstab}\label{thm:intro_stab}
    Let $(M,\tau)$ be a tracial von Neumann algebra. Fix $0\leq \eps\leq 1$ and assume $\{p_j\}_{j=1}^m$ is a family of projections in $M$ satisfying $\norm{1-\sum_{j=1}^mp_j}_2\leq \eps$.
    Then there exist projections $\{q_j\}_{j=1}^m$ in $M$ such that $\sum_{j=1}^m q_j = 1$ and $\sum_{j=1}^m\norm{p_j-q_j}_2^2\leq \bigO(\eps)$.
\end{restatable}

This stability theorem, and its stronger variant for contractions Theorem~\ref{thm:gen-stab}, improves on a previous result by Kim, Paulsen, and Schafhauser \cite[Lemma~3.5]{KPS18} (see also \cite{paddockthesis,Harris24}), whose result gives an exponential dependence on $m$ for the closeness between the operators and the rounded operators.

The proof of Theorem~\ref{thm:intro_stab} amounts to reducing the problem to \cite[Theorem~1.2]{dlS22orthog}, which is itself a strengthening and generalization of \cite[Theorem 5.2]{ji2020}). In de la Salle's result \cite[Theorem~1.2]{dlS22orthog}, the set-up is similar to ours with positive contractions instead of projections, and where the additional assumption that the positive contractions add to one (i.e., form a POVM) is imposed. In our case, the sum of the original operators may be larger than one, and our main task is thus to correct them in an appropriate way in order to be able to apply \cite[Theorem~1.2]{dlS22orthog}. We refer to Section~\ref{sec:stab} for the details. We also point out that the main difference between this approach and the approach from \cite{KPS18} is that we perturb all operators at the same time, whereas the proof of \cite[Lemma~3.5]{KPS18} proceeds inductively, perturbing the operators one by one, yielding exponential dependence, as compared to our result above and \cite[Theorem~1.2]{dlS22orthog}, where there is no dependence on the number of operators.

\,\\

\noindent \textbf{Related work. } To the best of our knowledge, the only other known gap-preserving reductions in the quantum setting have been shown by Culf and Mastel in \cite{CulfMastel24}. Our work and their work are complementary in many respects.

Stated in the language of this paper, Culf and Mastel show that for many NP-complete problems, there exists a constant $0<\overline{s}<1$ such that the associated $(1,\overline{s})$-gap$^*$ problem is undecidable. For instance, they lift Ji's gapless reduction for 3-Coloring \cite{ji2013binary} to a gap-preserving reduction. In contrast, the problem we consider in this paper is a classically easy problem, which cannot be captured by their result.

In terms of techniques, Culf and Mastel's work builds upon the framework developed by Mastel and Slofstra in \cite{MastelSlofstra24}.\footnote{In their work, Mastel and Slofstra perform a reduction from a gapped promise problem for synchronous games to a gapped promise problem for 3SAT albeit with a polynomially decreasing gap.} Therefore, their techniques are $^*$-algebraic as they work at the level of the algebras associated to Boolean Constraint System games. Moreover, they consider problems with constant answer size and perform their gap-preserving reductions from the RE-complete family of synchronous games with constant answer size from \cite{Dong24}. In comparison, our techniques work at the level of the strategies, yielding a more direct approach to showing gap-preserving reductions. Furthermore, we consider a problem with constant question size, which cannot be captured by the techniques of Culf and Mastel, and we perform our gap-preserving reduction from the RE-complete family of synchronous games with constant question size from \cite{NZ23}. Finally, we prove a new efficient stability theorem along the way, whereas Culf and Mastel rely on a known stability theorem from \cite{CVY23} which doesn't apply in our setting.

Another related work is that of Harris \cite{Harris24} who analyzes Ji's gapless reduction \cite{ji2013binary} in the approximate regime and, by using \cite[Lemma~3.5]{KPS18}, reaches a gap that decreases exponentially, which is not suitable for complexity-theoretic conclusions. Our stability result can be used to reduce the gap's exponential decrease to a polynomial decrease. It is an interesting open question whether his work combined with our stability theorem could give a gap-preserving reduction.
\,\\

\noindent \textbf{Future directions. }
Our work opens many new avenues for future progress within our proposed research program guided by the three questions \ref{introkeypoint1}--\ref{introkeypoint3} described above. Here we list a few specific open questions:
\begin{enumerate}
    \item \textbf{Complexity landscape of independent set games.} In Corollary \ref{cor:intro_hardness}, can we identify specific values of $t$ and $s$ for which the described problem is RE-hard? More precisely, how does the complexity of the $(1,s)$-gap$^*$ problem for $t$-independent set games vary with $s$ for a fixed $t$? We show that it can be RE-hard as well as trivial, but could there be for instance a range of $s$ where it is complete for some other known complexity class?

    \item \textbf{Improving our main theorem.} Could the gap in Theorem~\ref{thm:mainreduction} be improved to be independent of the number of questions $t$? This could imply that, similarly to synchronous games, independent set games are MIP$^*$-complete in the strong sense.\footnote{Recall that we say that a family of games is MIP$^*$-complete in the strong sense if the $(c,s)$-gap$^*$ associated to it is MIP$^*$-complete for all $0<s<c\leq 1$. Note that here we refer to the general class of independent set games which do not have a fixed question size. Indeed, for any fixed $t\in \N$, the class of $t$-independent set games cannot be MIP$^*$-complete in the strong sense, see Remark~\ref{rmk:trivialsoundness}.} However, we note that we show in Section~\ref{sec:sharpness} (see Remark~\ref{rmk:sharpness}) that this is impossible when using the construction of independent set games from this paper. Hence, resolving this question would involve discovering a new way to construct independent set games based on arbitrary synchronous games.

    \item \textbf{Gapped reductions in a different regime.} Beyond independent set games, it would be interesting to show reductions for $(c,s)$-gap$^*$ problems where $c\neq 1$. So far all the known gap-preserving reductions rely on stability theorems which are tied to the regime where $c=1$. This question is also motivated from the theory of hardness of approximation in the classical setting where reductions are often performed for $(c,s)$-gap problems where $c\neq 1$.
\end{enumerate} 
\,\\

\noindent \textbf{Organisation of the paper. } Besides the introduction, there are five other sections. In Section~\ref{sec:prelim} we collect some notation, definitions, and known results, and we describe the framework for our results. In Section~\ref{sec:analysis_perfect} we give a streamlined proof of the gapless reduction from \cite{MRV15}, since we will use it for our main result. In Section~\ref{sec:stab} we prove our stability Theorem~\ref{thm:intro_stab}, and in Section~\ref{sec:analysis_approx} we complete the proofs of our main results Theorem~\ref{thm:mainreduction} and Corollary~\ref{cor:intro_hardness}. Finally, in Section~\ref{sec:sharpness}, we provide an example showing that the bound in the main step in the proof of Theorem~\ref{thm:mainreduction} is sharp.
\,\\

\noindent \textbf{Acknowledgments. } We thank Eric Culf, Kieran Mastel, Connor Paddock and Yuming Zhao for helpful discussions. LM and TS are supported by the European Union via ERC grant QInteract (grant No 101078107) and VILLUM FONDEN via the QMATH Centre of Excellence (grant No 10059) and Villum Young Investigator (grant No 37532). PS is supported by an MSCA Fellowship (grant No 101111079) from the European Union. MV is supported by the NWO Vidi grant VI.Vidi.192.018 `Non-commutative harmonic analysis and rigidity of operator
algebras'.

\section{Preliminaries}\label{sec:prelim}

\subsection{von Neumann algebras}\label{ssec:vNa}

A \textit{von Neumann algebra} is a $^*$-subalgebra of the algebra $B(H)$ of bounded operators on some Hilbert space $H$ which is closed in the weak operator topology, i.e., the smallest topology such that all maps of the form $x\mapsto \langle x\xi,\eta\rangle$ for $\xi,\eta\in H$ are continuous. The famous von Neumann bicommutant theorem, see for instance \cite[Theorem~II.3.9]{Tak01}, states that $M\subset B(H)$ is a von Neumann algebra if and only if $M = M''$, where for a subset $S\subset B(H)$, $S'\coloneq \{y\in B(H)\mid xy=yx \text{ for all } x\in S\}$ denotes the commutant of $S$. For simplicity, we will assume that our von Neumann algebras are represented on a separable Hilbert space $H$, i.e., $H$ has a finite or countable orthonormal basis.

We will be concerned with \textit{finite} von Neumann algebras $M$, i.e., von Neumann algebras in which every isometry is a unitary. Equivalently, see for instance \cite[Theorem~V.2.4]{Tak01}, finite von Neumann algebras are exactly the von Neumann algebras which admit a positive linear functional $\tau:M\to \C$ with $\tau(1)=1$, which is
\begin{itemize}[topsep=2pt]
    \item faithful, i.e., for all $x\in M$: $\tau(x^*x)=0 \iff x=0$,
    \item normal, i.e., its restriction to the unit ball of $M$ is continuous in the weak operator topology, and
    \item tracial, i.e., for all $x,y\in M$: $\tau(xy)=\tau(yx)$.
\end{itemize}
We will refer to $\tau$ as a \textit{trace}, and to the pair $(M,\tau)$ as a \textit{tracial von Neumann algebra}. Given a tracial von Neumann algebra, we can associate a canonical trace-norm to it given by $\norm{x}_2 = \sqrt{\tau(x^*x)}$. 

\begin{ex}
    A first, and important, example of a tracial von Neumann algebra is the matrix algebra $\bM_n(\C)$ with its usual (normalized) trace. Readers interested solely in quantum strategies on finite-dimensional Hilbert spaces, can safely assume that all von Neumann algebras in the remainder of this paper are finite-dimensional, i.e., (direct sums of) matrix algebras.
\end{ex}

\subsection{Nonlocal games}\label{ssec:games}

In a (two-player) nonlocal game $\cG$, two players, usually called Alice and Bob, play against a verifier. The game $\cG$ is specified by four (finite) sets, consisting of question sets $Q$ and $Q'$, and answer sets $A$ and $A'$, together with a probability distribution $\pi$ on $Q\times Q'$ and a predicate $V: Q\times Q'\times A\times A'\to \{0,1\}$. By enlarging the question and answer sets if necessary, we will assume for simplicity that the question set and answer set are the same for both players. 
In fact, we will mostly deal with synchronous games in this paper, where this assumption is enforced. We thus denote a nonlocal game $\cG$ by a tuple $\cG=(Q,A,\pi,V)$. 
During the game, the verifier samples a question pair $(q,q')$ from $Q\times Q$ according to $\pi$, sends $q$ to Alice, $q'$ to Bob, and they respond with answers $a,a'\in A$ respectively. Alice and Bob win this round of the game if $V(q,q';a,a')=1$ and they lose otherwise. Importantly, while the setup of the game $\cG=(Q,A,\pi,V)$ is known to everyone, Alice and Bob cannot communicate during the game, and thus have no knowledge of each other's question and answer. Their goal is to come up with a strategy which maximizes their chances of winning. 

\vspace{8pt}

\noindent \textbf{Strategies for nonlocal games.} A classical deterministic strategy for $\cG$ consists of two functions $f:Q\to A$ and $f':Q\to A$ which Alice and Bob use to determine their answers. Using this strategy, the probability of winning the game $\cG$ is
\[
\omega(\cG, \{f,f'\}) = \sum_{(q,q')\in Q\times Q} \pi(q,q')V(q,q';f(q), f'(q')),
\]
and the classical value $\omega(\cG)$ (also denoted $\omega_{loc}(\cG)$) of the game $\cG$ is the maximum of these probabilities over all possible such strategies. More generally, a classical strategy could also make use of shared randomness between the players. It is easy to see, however, that strategies using shared randomness do not provide an advantage over deterministic strategies. 

On the other hand, a \textit{quantum strategy} for $\cG$ allows the players to determine their answers based on joint measurements on a shared finite-dimensional entangled state. This is modeled using so-called Positive Operator Valued Measurements (POVM), i.e., sets of positive operators on a Hilbert space which sum to the identity. However, it is well-known that by potentially enlarging the Hilbert spaces they act on, it is possible to consider Projective Value Measurements (PVMs) instead, i.e., sets of projections on a Hilbert space which sum to the identity. More precisely, a quantum strategy $\cS$ for a nonlocal game $\cG=(Q,A,\pi,V)$ consists of the following:
\begin{enumerate}
    \item Two finite dimensional Hilbert spaces $H, H'$;
    \item For every $q\in Q$, a PVM $\{M_q^a\}_{a\in A}$ on $H$ and a PVM $\{N_{q}^{a}\}_{a\in A}$ on $H'$;
    \item A quantum state $\ket{\psi}\in H\otimes H'$.
\end{enumerate}
Here we think of the $M_q^a$ as the operators from Alice's measurement, and the $N_q^a$ as the operators from Bob's measurement. Using this quantum strategy, the probability that the players measure and answer $a,a'\in A$  when given the questions $q,q'\in Q$ respectively is 
\[
p(a,a'\lvert q,q')=\bra{\psi} M_q^a \otimes N_{q'}^{a'} \ket{\psi}.
\]
In particular, the probability of winning the game $\cG$ using the strategy $\cS$ is
\[
\omega^*(\cG, \cS) = \sum_{(q,q')\in Q\times Q} \sum_{(a,a')\in A \times A} \pi(q,q')V(q,q';a,a')\bra{\psi} M_q^a \otimes N_{q'}^{a'} \ket{\psi},
\]
and the quantum value $\omega^*(\cG)$ of the game $\cG$ is the supremum of $\omega^*(\cG, \cS)$ over all possible quantum strategies $\cS$. We note that this quantum value is sometimes denoted by $\omega_q(\cG)$, but we stick to the above notation $\omega^*$ in this paper. We also point out that, unlike the case of classical strategies, there does not necessarily exist a quantum strategy realizing this supremum, see \cite{Slo19}.

One more common class of strategies for nonlocal games are the \textit{quantum commuting} strategies. A quantum commuting strategy $\cS$ for $\cG = (Q,A,\pi,V)$ consists of the following:
\begin{enumerate}
    \item A Hilbert space $H$;
    \item For every $q\in Q$, two PVMs $\{M_q^a\}_{a\in A}$ and $\{N_{q}^{a}\}_{a\in A}$ on $H$ satisfying $M_q^aN_{q'}^{a'} = N_{q'}^{a'}M_q^a$ for all $q,q'\in Q, a,a'\in A$;
    \item A quantum state $\ket{\psi}\in H$.
\end{enumerate}
One can again consider the probability that the players measure and answer $a,a'\in A$  when given the questions $q,q'\in Q$ respectively, which is given by $p(a,a'\lvert q,q')=\bra{\psi} M_q^a N_{q'}^{a'} \ket{\psi}$. The quantum commuting value $\omega_{qc}(\cG)$ of $\cG$ is then the supremum of the winning probabilities over all quantum commuting strategies. Since we allow for arbitrary Hilbert spaces, it is not too difficult to show that this supremum is in fact attained by some quantum commuting strategy. Also, we note that the quantum commuting strategies contain the quantum strategies.

\vspace{8pt}

\noindent \textbf{Synchronous games and synchronous strategies.} A nonlocal game $\cG$ is a \emph{synchronous game} if it has just one question set $Q$ and one answer set $A$ for both Alice and Bob, and the predicate satisfies the following: for any $q\in Q$, $V(q,q;a,a')=1$ if and only if $a=a'$. In other words, given the same question, Alice and Bob are required to give the same answer in order to win the game.

Given a synchronous game $\cG$, a strategy for $\cG$ is called \textit{synchronous} if the corresponding correlation satisfies $p(a,a'\lvert q,q) = 0$ if $a\neq a'$ for any $q\in Q$. The synchronous quantum value $\omega^*_s(\cG)$ (or $\omega^s_q(\cG)$) of the synchronous game $\cG$ is the supremum of the winning probabilities over all synchronous quantum strategies. Similarly, one can consider the synchronous quantum commuting value $\omega^s_{qc}(\cG)$ of the synchronous game $\cG$, as the supremum of the winning probabilities over all synchronous quantum commuting strategies.

A natural question is whether one can characterize the synchronous strategies among general quantum (commuting) strategies. In fact, since different strategies can yield the same correlation, it is more natural to phrase this question in terms of the correlation sets $C_q$ and $C_{qc}$ which consist of the correlations $p(a,a'\lvert q,q')$ arising from quantum and quantum commuting strategies respectively, and their synchronous versions $C_q^s \subset C_q$ and $C_{qc}^s\subset C_{qc}$ for which we restrict to synchronous strategies. It turns out that there is an elegant answer established in \cite{PSSTW16}, which we describe below. We note that the results in \cite{PSSTW16} are formulated in terms of $C^*$-algebras, whereas we choose to present them in terms of (tracial) von Neumann algebras, since this is more convenient for us in the remainder of the paper. Nevertheless, we point out that both formulations are equivalent in this case, merely by changing between the $C^*$-algebra and the von Neumann algebra generated by a given collection of PVMs.

\begin{thm}[{\cite[Corollary~5.6]{PSSTW16}}]\label{thm:synvNa}
    A correlation $p(a,a'\lvert q,q')$ belongs to $C_{qc}^s$ (resp. $C_q^s$) if and only if there exists a tracial von Neumann algebra (resp. finite-dimensional von Neumann algebra) $M$ with trace $\tau$ and a generating family of projections $\{P_q^a\mid q\in Q, a\in A\}$ satisfying $\sum_{a\in A} P_q^a = 1$ for every $q\in Q$ such that 
    \[
    \forall q,q'\in Q, a,a'\in A:\quad p(a,a'\lvert q,q') = \tau(P_q^a P_{q'}^{a'}).
    \]
\end{thm}

Thanks to the above theorem, when computing the synchronous quantum (commuting) value of a game, one can work with collections of PVMs in tracial von Neumann algebras instead of dealing with actual strategies. We will mostly take this approach in the remainder of the paper, since a lot of our results boil down to perturbing the projections in approximate PVMs to get genuine PVMs. 

\begin{rmk}
    One can also characterize the correlations belonging to other correlation sets, such as $C_{loc}^{s}$ \cite{PSSTW16}, and $C_{qa}^{s}$ and $C_{qs}^{s}$ \cite{KPS18} in terms of traces. Our results can be interpreted accordingly for these correlations sets.
\end{rmk}

Given a synchronous game, another natural question to ask is how the quantum (commuting) value and the synchronous quantum (commuting) value are related. The following result due to de la Salle and Marrakchi \cite{dlSM23} (and established by Vidick in the finite-dimensional setting in \cite{Vidick_2022}) is useful in translating between the two, see also \cite{paddock22, lin23}. In order to state the theorem formally, we call a probability distribution $\pi$ on $Q\times Q$ a \textit{$C$-diagonally dominant distribution} if 
\[
\pi(q,q) \geq C \cdot \max\left\{\sum_{q'\in Q} \pi(q,q'), \sum_{q'\in Q} \pi(q',q)\right\},
\]
for all $q\in Q$. For instance, if the distribution $\pi$ is uniform then we have $C=1/\abs{Q}$ since:
\begin{align*}
    \pi(q,q)=\frac{1}{\abs{Q}^2} = \frac{1}{\abs{Q}} \sum_{q' \in Q} \pi(q,q')=\frac{1}{\abs{Q}} \sum_{q' \in Q} \pi(q',q).
\end{align*}

\begin{thm}[{\cite{dlSM23}}]\label{thm:dlSM23}
    Suppose $\mathcal{G}$ is a synchronous game with a $C$-diagonally dominant question distribution. For $\mathrm{x} = q,qc$, if $\omega_{\mathrm{x}}(\mathcal{G}) \geq 1-\varepsilon$, then $\omega^s_{\mathrm{x}}(\mathcal{G}) \geq 1-\bigO(\left(\frac{\varepsilon}{C}\right)^{\frac{1}{4}})$.
\end{thm}

For nonlocal games with a uniform question distribution, it follows from the above that the difference between the quantum (commuting) value and the synchronous quantum (commuting) value depends on the number of questions. However, we note that for any game $\cG$ and any constant $C\in [0,1/2)$, we can easily construct a $C$-diagonally dominant version of the game. This natural construction appears for example in \cite[Definition 2.4]{VZ25} under a different name for symmetric synchronous games, and similar ideas have appeared in the literature before, see for instance \cite{CulfMastel24}. 

\begin{defn}
    Let $\cG=(Q,A,\pi,V)$ be a nonlocal game and let $C\in[0,1/2)$. The \textit{$C$-diagonally weighted version of $\cG$} is the nonlocal game $\cG'=(X,A,\pi',V)$ with
    \begin{equation*}
        \pi'(x,y)=\frac{C}{2}\left(\sum_{z\in X}\pi(x,z)+\pi(z,x)\right)\delta_{x,y}+(1-C)\pi(x,y).
    \end{equation*}
\end{defn}

In a $C$-diagonally weighted version of a game, we have added a total weight of $C$ to the diagonal and it is clear that a $C$-diagonally weighted version of a game is $C/2$-diagonally dominant. 

\vspace{8pt}

\noindent \textbf{Independent set games.} Finally, we introduce the specific class of games we are mainly interested in, namely independent set games, which were first implicitly considered by Man\v{c}inska and Roberson in \cite{MR12}, and which were formally defined by Man\v{c}inska, Roberson, and Varvitsiotis in \cite{MRV15}:

\begin{defn}\label{def:indsetgame}
    Let $X$ be a simple graph with vertex set $\mathcal{V}(X)$ and let $t$ be a positive integer. Then the \textit{$t$-independent set game for $X$} has question set $[t]=\{1, 2, \ldots, t\}$, answer set $\mathcal{V}(X)$, and predicate $V: [t]\times [t] \times \mathcal{V}(X)\times \mathcal{V}(X) \to \{0,1\}$ satisfying $V(i,j;u,v)=1$ if and only if the following is satisfied:
    \begin{itemize}
    \item if $i=j$, then $u=v$, and
    \item if $i\neq j$, then $u$ and $v$ are distinct, non-adjacent vertices.
\end{itemize}
    We consider two different probability distributions for the $t$-independent set game. The probability distribution $\pi_u$ for the \emph{uniform} $t$-independent set game is the uniform distribution on $[t]\times [t]$. The probability distribution $\pi_d$ for the \emph{diagonally weighted} $t$-independent set game is given by
    \begin{equation*}
        \pi_d(i,j)=\frac{1}{2t}\delta_{i,j}+\frac{1}{2t^2}.
    \end{equation*}
\end{defn}

We will typically work with diagonally weighted independent set games, as these are $1/2$-diagonally dominant and therefore behave better for our purposes. We denote the diagonally weighted $t$-independent set game for $X$ by $(X,t)$. 

Every synchronous game has a game graph associated to it as in the following definition. Using this, we will later associate to a synchronous game $\cG = (Q,A,\pi,V)$ the $\abs{Q}$-independent set game for its game graph.

\begin{defn}[{\cite{MRV15}}]\label{def:gamegraph}
Let $\cG=(Q,A,\pi,V)$ be a synchronous game. The \textit{game graph} of $\cG$, denoted $X(\cG)$, is the undirected graph with vertex set $Q \times A$, and where $(q,a)$ is adjacent to $(q',a')$ if and only if  $V(q,q';a,a')=0$ or $V(q',q;a',a)=0$.
\end{defn}

For later reference, we explicitly compute the value of the diagonally weighted $t$-independent set game for $X(\cG)$ when using a synchronous strategy. Let $\cS$ be a synchronous strategy whose associated correlation is given by a collection of PVMs $\{P_i^{(q,a)}\}_{(q,a)\in \mathcal{V}(X)}$ indexed by $i\in [t]$ in a tracial von Neumann algebra $(M,\tau)$ as in Theorem~\ref{thm:synvNa}. Then, it follows immediately from the definition that the losing probability of this strategy is
\begin{align}\label{eq:loss_indep}
\begin{split}
    1-\omega^*((X(\cG),t),\cS)=\frac{1}{2t^2}\sum_{\substack{i \neq j,\\ (q,a)}} \tau\left(P_i^{(q,a)} P_j^{(q,a)}\right)+ \frac{1}{2t^2}\sum_{\substack{i\neq j, \\ V(q,q';a,a')=0\, \mathrm{or} \\ V(q',q;a',a)=0}} \tau\left(P_i^{(q,a)} P_j^{(q',a')}\right).
\end{split}
\end{align}

\subsection{Complexity theory}\label{ssec:complexity}

In this work, the computational problems we consider are so-called gapped promise problems.

\begin{defn}
    Let $0<s\leq 1$ be a constant. The \textit{gapped promise problem with completeness 1 and soundness $s$} for a family of games $C$ is the following problem. Given a nonlocal game $\cG \in C$, decide whether:
    \begin{itemize}
        \item $\omega^*(\cG)=1$ or
        \item $\omega^*(\cG)<s$,
    \end{itemize}
    given the promise that one of the two cases holds. We denote this problem the $(1,s)$-\textit{gap}$^*$ \textit{problem for} $C$.
\end{defn}

The seminal result MIP$^*$=RE of Ji, Natarajan, Vidick, Wright, and Yuen \cite{ji2020} shows that for any constant $0<s\leq 1$ there exists a family of games $C$ for which the $(1,s)$-gap$^*$ problem is RE-complete, where RE is the class of recursively enumerable languages. The family of games they construct consists of synchronous games which are \emph{succinctly presented}, see Definition \ref{def:succinctrep}. 

\begin{rmk}
    In this paper, we consider the question and answer sets of a game to be represented in binary. For a game $\cG = (Q,A,\pi,V)$, we have $Q=\{0,1\}^l$ and $A=\{0,1\}^m$ for some $l,m\in \N$. We call $l,m$ the \emph{question size} and the \emph{answer size} of the game respectively. 
\end{rmk}

\begin{defn}\label{def:succinctrep}
    A family of games $C=\{\cG_n\}_n$, where $\cG_n=(Q_n,A_n,\pi_n,V_n)$, is said to be \emph{succinctly presented} if there exist a probabilistic Turing machine $S$ and a Turing machine $V$ such that 
    \begin{itemize}
        \item on input $n$, $S$ samples a pair $(q,q')\in Q\times Q$ with probability $\pi_n(q,q')$,
        \item on input $(n,q,q',a,a') \in \N \times Q \times Q \times A \times A$, $V$ outputs $V_n(q,q';a,a')$.
    \end{itemize}
    Furthermore, we call the succinctly presented family of games $C$ \emph{efficient} if the question and answer sizes of $\cG_n$ scale as $\bigO(\mathrm{poly}(n))$ and the above Turing machines $S,V$ can be chosen to run in time $\mathrm{poly}(n)$.
\end{defn}

\begin{rmk}
    In this paper, we only consider succinctly presented families of games which are also efficient. For the sake of brevity, we therefore omit the term ``efficient'' for succinctly presented families of games.
\end{rmk}

The reason for considering succinctly presented family of games is that we are interested in the complexity of computing the value of games where sampling the questions and verifying the answers can be done efficiently. We therefore consider the most general setting where the verifier in the game is efficient while the instance sizes may grow exponentially in the family parameter.

We can now formally state the result of MIP$^*$=RE as follows.

\begin{thm}(\cite{ji2020})\label{thm:mip*=re}
    For every constant $0<s<1$, there exists a succinctly presented family $C=\{\cG_n\}_n$ of synchronous games whose question and answer sizes grow polynomially in $n$, such that the $(1,s)$-gap promise problem for $C$ is $\mathrm{RE}$-complete.
\end{thm}

As discussed in the introduction, the above result is not sufficient for our application and we need the following refinement proved by Natarajan and Zhang in \cite{NZ23}.

\begin{thm}(\cite{NZ23})\label{thm:NZ23}
    For every constant $0<s<1$, there exists a succinctly presented family $C=\{\cG_n\}_n$ of synchronous games, with constant question size and with answer size scaling as $\mathrm{polylog}(n)$, such that the $(1,s)$-gap promise problem for $C$ is $\mathrm{RE}$-complete. Moreover, the question distribution of each game in $C$ is uniform.
\end{thm}

\begin{rmk}\label{rmk:NZ23}
    In Theorem \ref{thm:NZ23}, the smaller the soundness parameter $s$ is, the bigger the constant question size will be. Indeed, to decrease the soundness parameter, parallel repetition is used in \cite{NZ23}, which increases the question size. The exact dependence of the question size on the soundness parameter will not be necessary for our application. 
\end{rmk}

We can now define gap-preserving reductions between promise problems. Intuitively, such a reduction consists of a map from instances of $C$ to instances of $\overline{C}$ which satisfies completeness and soundness properties. However, as it is important to work with succinctly presented problems, we need to phrase the definition formally using Turing machines which implement the map between instances.

\begin{defn}\label{def:reduction}
    Let $0<s,\overline{s} \leq 1$ be constants, and let $C=\{\cG_n\}_n$ and $\overline{C}=\{\overline{\cG}_{m}\}_{m}$ be succinctly presented families of games where $\cG_n=(Q_n,A_n,\pi_n,V_n)$ and $\overline{\cG}_{m}=(\overline{Q}_{m},\overline{A}_{m},\overline{\pi}_{m},\overline{V}_{m})$. Let $S,V$ be any Turing machines associated to $C$ as in Definition \ref{def:succinctrep}. A \emph{reduction} from the $(1,s)$-gap problem for $C$ to the $(1,\overline{s})$-gap problem for $\overline{C}$ consists of a probabilistic Turing machine $\overline{S}$ and Turing machine $\overline{V}$ satisfying the following conditions:
    \begin{itemize}
        \item $\overline{S}$ has oracle access to $S$ and on input $n$ outputs $\overline{n}\in \N$ as well as $(\overline{q},\overline{q}')\in \overline{Q}_{\overline{n}}\times \overline{Q}_{\overline{n}} $ with probability $\overline{\pi}_{\overline{n}}(\overline{q},\overline{q}')$,
        \item $\overline{V}$ has oracle access to $V$ and on input $(n,\overline{q},\overline{q}',\overline{a},\overline{a}')\in \N \times \overline{Q}_{\overline{n}}\times \overline{Q}_{\overline{n}} \times \overline{A}_{\overline{n}} \times \overline{A}_{\overline{n}}$ outputs $\overline{V}_{\overline{n}}(\overline{q},\overline{q}',\overline{a},\overline{a}')$,
        \item $\overline{S}$ and $\overline{V}$ run in time $\mathrm{poly}(n)$,
        \item (completeness) if $\omega^*(\cG_n)=1$, then $\omega^*(\overline{\cG}_{\overline{n}})=1$,
        \item (soundness) if $\omega^*(\cG_n)<s$, then $\omega^*(\overline{\cG}_{\overline{n}})<\overline{s}$.
    \end{itemize}
\end{defn} 

The intuition behind this definition of reduction is the following: a verifier playing a game $\cG\in C$ could instead efficiently play the associated game $\overline{\cG}\in \overline{C}$ as it can efficiently convert the Turing machines $S,V$ for $C$ to Turing machines $\overline{S},\overline{V}$ for $\overline{C}$. Due to the completeness and soundness properties, deciding the gapped promise problem for the family $\overline{C}$ must be as hard as deciding the gapped promise problem for the family $C$. In particular, we have the following:

\begin{lem}\label{lem:REred}
    If there is a reduction from the $(1,s)$-gap problem for $C$ to the $(1,\overline{s})$-gap problem for $\overline{C}$ and deciding the $(1,s)$-gap problem for $C$ is $\mathrm{RE}$-hard, then deciding the $(1,\overline{s})$-gap problem for $\overline{C}$ is also $\mathrm{RE}$-hard.
\end{lem}

\begin{proof}
    This follows immediately from the completeness and soundness properties of a reduction. 
\end{proof}

\begin{rmk}\label{rmk:liftinggapless}
    A gapless reduction is a reduction as in Definition \ref{def:reduction} where both $s=\overline{s}=1$. As mentioned in the introduction, there are a number of known examples of gapless reductions in the literature \cite{ji2013binary,MRV15,Harris24a,Atserias}. It is usually not too complicated to lift a gapless reduction to a reduction where the resulting gap $\overline{s}$ depends on the family parameter $n$, e.g., $s<1$ is constant but $\overline{s}=1-1/\bigO(\mathrm{poly}(n))$. For instance, this lifting can be done for the gapless reduction of \cite{MRV15} in a relatively straightforward manner. However, such a reduction is not meaningful from a complexity theoretic point of view. Removing the dependence on $n$ from the gap is a challenging and subtle task, which is the main result of our work. Indeed, Theorem \ref{thm:mainreduction} can be seen as lifting the gapless reduction of \cite{MRV15} to an actual gap-preserving reduction, where $0<s,\overline{s}<1$ are both constants.
\end{rmk}

\section{Perfect regime}\label{sec:analysis_perfect}

In this section, we review the proof of the following theorem due to Man\v{c}inska, Roberson, and Varvitsiotis \cite{MRV15} in the framework from Section~\ref{ssec:games}. Our main result Theorem~\ref{thm:mainreduction} is a generalization of this result to the gapped case.

\begin{thm}[{\cite{MRV15}}]\label{thm:MRV15}
    Let $\cG=(Q,A,\pi,V)$ be a synchronous game, $X(\cG)$ be its corresponding game graph, and $t=\abs{Q}$ be the number of questions. Then $\cG$ has a perfect quantum strategy if and only if the independent set game $(X(\cG),t)$ has a perfect quantum strategy..
 \end{thm}

\begin{proof}
Since $\cG$ is a synchronous game, any perfect strategy is necessarily synchronous. Hence, let $\cS$ be a perfect synchronous strategy for the game $\cG$ given by PVMs $\{P_q^a\}_{a\in A}$ for every $q\in Q$ on a tracial von Neumann algebra $(M,\tau)$. We can use this same strategy to play the independent set game $(X(\cG),t)$. Indeed, as the independent set is of size $t=\abs{Q}$, we can label questions in $[t]$ by elements $q\in Q$. When the provers receive question $q$ they measure the PVM $\{P_q^a\}_{a\in A}$ and answer with vertex $(q,a)\in Q \times A $ upon getting the measurement outcome $a\in A$. The losing probability of this strategy $\overline{\cS}$ for the independent set game $(X(\cG),t)$ is given by replacing the operators $P_q^{(q,a)}$ with $P_q^a$ in equation \eqref{eq:loss_indep}. With this replacement, the first sum in equation \eqref{eq:loss_indep} is zero by definition and thus we get: 
\begin{align*}
    1-\omega^*((X(\cG),t),\overline{\cS})= \frac{1}{t^2}\sum_{\substack{q\neq q', \\ V(q,q';a,a')=0\, \mathrm{or} \\ V( q',q;a',a)=0}} \tau(P_q^{a} P_{q'}^{a'})=0.
\end{align*}
In other words, $\overline{\cS}$ is a perfect strategy for $(X(\cG),t)$. This concludes the first direction of the proof. 

Conversely, let $\cS'$ be a perfect synchronous strategy for the independent set game $(X(\cG),t)$ given by PVMs $\{P_i^{(q,a)}\}_{(q,a)\in Q\times A}$ for every $i\in [t]$ on a tracial von Neumann algebra $(M,\tau)$. Define the operators
\begin{align*}
    P_q^a\coloneqq \sum_i P_i^{(q,a)}.
\end{align*}
We show that for every $q\in Q$ the set of operators $\{P_q^a\}_{a \in A}$ forms a PVM. Firstly, we check that for all $q\in Q, a\in A$, the operators $P_q^a$ are projections. They are clearly self-adjoint, and furthermore
\begin{align*}
    P_q^a P_q^{a} = \sum_{i,j} P_i^{(q,a)} P_j^{(q,a)} = \sum_i P_i^{(q,a)} + \sum_{i\neq j} P_i^{(q,a)} P_j^{(q,a)} = \sum_i P_i^{(q,a)} = P_q^a,
\end{align*}
where we use that since the strategy $\cS'$ is perfect, equation~\eqref{eq:loss_indep} implies that $P_i^{(q,a)} P_j^{(q,a)}=0$ for all $i\neq j$.\footnote{We additionally make use of the fact that for $A,B$ positive, $AB=0$ if and only if $\tau(AB)=0$. The forward direction is clear. For the other direction, write $A=S^*S$, $B=T^*T$, then define $X=ST^*$ and notice that $0=\tau(AB)=\tau(X^*X)=\norm{X}_\tau^2$ which implies that $X=0$ and thus $AB=S^*XT=0$.} Next, we check that for every $q\in Q$, the operators $P_q^a$ and $P_q^{a'}$ are orthogonal for $a\neq a'$:
\begin{align*}
    P_q^a P_q^{a'} = \sum_{i,j} P_i^{(q,a)} P_j^{(q,a')} = \sum_i P_i^{(q,a)} P_i^{(q,a')} + \sum_{i\neq j} P_i^{(q,a)} P_j^{(q,a')} =0,
\end{align*}
where we use that $\{P_i^{(q,a)}\}_{(q,a)}$ forms a PVM to cancel the first sum. For the second sum, we notice that $V(q,q;a,a')=0$ by synchronicity and thus we conclude from equation~\eqref{eq:loss_indep} as above that $P_i^{(q,a)} P_j^{(q,a')}=0$ for all $i \neq j$. It remains to show that for every $q\in Q$ the operators in $\{P_q^a\}_{a \in A}$ sum to the identity. Note that as the operators in $\{P_q^a\}_{a\in A}$ are mutually orthogonal projections, we have $\sum_a P_q^a \leq 1$. Since $\{P_i^{(q,a)}\}_{(q,a)\in Q\times A}$ is a PVM for every $i$, we thus get
\begin{align*}
    \abs{Q} = \sum_i \sum_{q,a} P_i^{(q,a)} = \sum_q \sum_a P_q^a \leq \sum_q 1 = \abs{Q}, 
\end{align*}
which implies that $\sum_a P_q^a=1$.

We now show that the synchronous strategy $\cS$ given by the PVMs $\{P_q^a\}_{a\in A}$ for every $q\in Q$ on the tracial von Neumann algebra $(M,\tau)$ is a perfect strategy for the synchronous game $\cG$. Indeed, the losing probability of the strategy $\cS$ is given by
\begin{align*}
    1-\omega^*(\cG,\cS)&=\sum_{\substack{q,q',a,a'\\ V(q,q';a,a')=0}} \pi(q,q') \tau(P_q^a P_{q'}^{a'})\\ &= \sum_{\substack{q,q',a,a'\\ V(q,q';a,a')=0}} \sum_{i,j} \pi(q,q') \tau(P_i^{(q,a)} P_j^{(q',a')})\\ &= \sum_{i,j} \sum_{(q,a) \sim_{X(\cG)} (q',a')} \pi(q,q') \tau(P_i^{(q,a)} P_j^{(q',a')})\\
    &\leq \sum_{i} \sum_{(q,a)\neq (q',a')} \pi(q,q') \tau(P_i^{(q,a)} P_i^{(q',a')})\\ &\qquad + \sum_{i\neq j} \sum_{(q,a) \sim_{X(\cG)} (q',a')} \pi(q,q') \tau(P_i^{(q,a)} P_j^{(q',a')})\\
    &= 0,
\end{align*}
where the last inequality follows from equation~\eqref{eq:loss_indep} together with the fact that the strategy $\cS'$ is perfect. Hence $\cS$ is a perfect strategy for $\cG$, finishing the proof of the theorem.
\end{proof}

\begin{rmk}\label{rmk:MRV15}
    Since the quantum value of a game is not necessarily attained, the theorem as stated does not imply that $\omega^*(\cG) = 1$ if and only if $\omega^*(X(\cG),t) = 1$. Nevertheless, one can prove that this is actually the case, by using that $\omega^*(\cG) = 1$ if and only if for every $\eps>0$, there exists a quantum strategy $\cS$ for $\cG$ with $\omega^*(\cG,\cS) > 1-\eps$. The ``only if'' of the above statement then follows in exactly the same way as in the proof of Theorem~\ref{thm:MRV15}, since such a strategy $\cS$ can be used for the independent set game $(X(\cG),t)$ yielding quantum value $\omega^*((X(\cG),t),\cS) > 1-\eps$. For the ``if''-statement, one needs a stability theorem to round the projections of a given strategy for $(X(\cG),t)$ in order to get a genuine strategy for $\cG$. This is possible using stability theorems available in the literature, such as \cite[Lemma~3.5]{KPS18}. However, it also follows from our more general Theorem~\ref{thm:mainreduction}, so we won't go into the details here.
\end{rmk}

\section{A stability theorem}\label{sec:stab}
In this section, we prove our main stability result, Theorem \ref{thm:intro_stab}, and its more general version, Theorem \ref{thm:gen-stab}. This result is the cornerstone for the proof of our main Theorem \ref{thm:mainreduction}. In order to make the results in this section applicable to both quantum strategies as well as quantum commuting strategies, the results and proofs are formulated in terms of a general tracial von Neumann algebra $(M,\tau)$. However, we point out that there is no harm in assuming $M$ to be finite-dimensional if one is mainly interested in quantum strategies. 

\begin{thm}\label{thm:gen-stab}
    Let $(M,\tau)$ be a tracial von Neumann algebra. Fix $0\leq \eps, \delta \leq 1$ and assume $\{a_j\}_{j=1}^m$ is a family of positive operators in $M$ satisfying the following:
    \begin{enumerate}[label=(\alph*)]
        \item for every $1\leq j\leq m$: $a_j\leq 1$,
        \item $\norm{1-\sum_{j=1}^ma_j}_2\leq \eps$,
        \item $\left|1-\tau(\sum_{j=1}^ma_j^2)\right|\leq \delta$.
    \end{enumerate}
    Then there exists a PVM $\{q_j\}_{j=1}^m$ in $M$ such that $\sum_{j=1}^m\norm{a_j-q_j}_2^2\leq \bigO(\eps+\delta)$.
\end{thm}

Our proof revolves around reducing our setting to the setting of \cite[Theorem 1.2]{dlS22orthog}. To make clear how the current assumptions differ from the ones in \cite[Theorem 1.2]{dlS22orthog}, we first recall this theorem.
\begin{thm*}[{\cite[Theorem 1.2]{dlS22orthog}}]
    Let $M$ be a von Neumann algebra with a normal state $\phi$, and let $\{a_j\}_{j=1}^m$ be a POVM in $M$ such that $\phi(\sum_{j=1}^ma_j^2)> 1-\eps'$. Then there exists a PVM $\{q_j\}_{j=1}^m$ in $M$ such that $\phi(\sum_{j=1}^m|a_j-p_j|^2)<9\eps'$.
\end{thm*}
The condition that $\phi(\sum_{j=1}^ma_j^2)>1-\eps$ is also present in our theorem as assumption (c). However, in our Theorem~\ref{thm:gen-stab}, the family $\{a_j\}_{j=1}^m$ no longer has to sum to one exactly, which is the condition for being a POVM. Instead, assumption (b) requires that the family sums to 1 approximately (in trace-norm). Hence, our task is to modify the family $\{a_j\}_{j=1}^m$ in such a way that it does sum to one. We do this in two steps. First, we add an operator $a_0$ to the family to ensure that it sums to \emph{at least} one. Next, we rescale all operators $a_j$ to make the family sum to exactly one, obtaining a POVM and allowing us to use \cite[Theorem 1.2]{dlS22orthog}. In each step, we carefully keep track of the accumulated error, which yields the desired result.

\begin{proof}[Proof of Theorem~\ref{thm:gen-stab}]
    Define the self-adjoint operator
    \begin{equation*}
        x=1-\sum_{j=1}^ma_j
    \end{equation*}
    and let $a_0=x_+$ be the positive part of $x$. Defining 
    \begin{equation*}
        \rho=\sum_{j=0}^ma_j,
    \end{equation*}
    we then have that $\rho\geq 1$. In particular, we have obtained a family of positive operators that sums to at least one. Moreover,
    \begin{equation*}
        \norm{a_0}_2^2 = \tau\left(a_0^2\right) \leq \tau\left(a_0^2+x_-^2\right) = \tau\left(x^2\right) = \norm{1-\sum_{j=1}^ma_j}_2^2 \leq \eps^2,
    \end{equation*}
    where in the third step we use that the product $y_+y_-$ of the positive and negative part of any self-adjoint operator $y$ equals zero. Consequently, $\norm{a_0}_2\leq \eps$. Hence, defining for $1\leq j\leq m$,
    \begin{equation*}
        b_j=\begin{cases}
            a_0+a_1 & \text{ if }j=1\\
            a_j & \text{ otherwise}
        \end{cases}
    \end{equation*}
    we deduce that 
    \begin{equation*}
        \left(\sum_{j=1}^m\norm{a_j-b_j}_2^2\right)^{1/2}=\norm{a_0}_2\leq \eps.
    \end{equation*}
    Furthermore, by construction, 
    \begin{equation*}
        \sum_{j=1}^mb_j = \sum_{j=0}^m a_j = \rho \geq 1.
    \end{equation*}
    Lastly, 
    \begin{equation}\label{eq:rhominusone}
        \norm{\rho-1}_2^2=\tau\left(x_-^2\right)
        \leq \tau\left(x_-^2+x_+^2\right)
        =\tau\left(x^2\right)
        =\norm{-1+\sum_{j=1}^ma_j}^2
        \leq \epsilon^2,
    \end{equation}
    so $\norm{\rho-1}_2\leq \epsilon$. 

    In the next step, we aim to turn the family $\{b_j\}_{j=1}^m$ into a POVM. For this, note that $\rho$ is invertible since $\rho\geq 1$, and thus we can define 
    \[
    c_j=\rhomh b_j\rhomh.
    \]
    Clearly, $\{c_j\}_{j=1}^m$ is a POVM, and we proceed with showing that the $c_j$ are close to the $b_j$ in the appropriate sense. Firstly, we have
    \begin{align*}
        \sum_{j=1}^m\norm{b_j-c_j}_2^2 &= \sum_{j=1}^m\tau\left(b_j^2-2b_j\rhomh b_j\rhomh + \rhomh b_j \rhomo b_j\rhomh \right)\\
        &= \sum_{j=1}^m\tau\left(b_j^2-2b_j\rhomh b_j\rhomh + b_j \rhomo b_j\rhomo \right)
    \end{align*}
    Note that since $\rho\geq 1$, we have $\rhomh \leq 1$ and thus also $\rhomo b_j \rhomo\leq \rhomh b_j\rhomh$. In particular, $\tau(b_j \rhomo b_j\rhomo) \leq\tau(b_j\rhomh b_j\rhomh)$, where we use the fact that $\tau(xy)\leq \tau(xz)$ whenever $x\geq 0$ and $y\leq z$. Using this, we find
    \begin{align*}
        \sum_{j=1}^m\norm{b_j-c_j}_2^2&\leq \sum_{j=1}^m\tau\left(b_j^2 - b_j \rhomh b_j\rhomh \right)\\
        &=\sum_{j=1}^m\tau\left(b_j^2(1-\rhomh)+b_j(1-\rhomh)b_j\rhomh\right)\\
        &\leq 2\sum_{j=1}^m\tau\left(b_j^2(1-\rhomh)\right),
    \end{align*}
    where for the last inequality, we used the fact that $\rhomh \leq 1$ to deduce similar to the above that $\tau(b_j(1-\rhomh)b_j\rhomh) \leq \tau(b_j(1-\rhomh)b_j) = \tau(b_j^2(1-\rhomh))$.
    
    Since $b_j\leq 1$ for all $2\leq j\leq m$ by assumption and $b_1\leq 2$ by construction, we also have that $b_j^2\leq 2b_j$ for every $1\leq j\leq m$. Hence,
    \begin{equation}\label{eq:c-b-mod-error-est}
        \sum_{j=1}^m\norm{b_j-c_j}_2^2\leq 4\sum_{j=1}^m\tau\left(b_j(1-\rhomh)\right)=4\tau\left(\rho(1-\rhomh)\right)\leq 4\tau\left(\rho-1\right)\leq 4\eps,
    \end{equation}
    where we use the estimate $\rho^{1/2}\geq 1$ in the penultimate step and the Cauchy-Schwarz inequality together with \eqref{eq:rhominusone} in the final step. 

    To apply \cite[Theorem 1.2]{dlS22orthog}, we need to estimate $\sum_{j=1}^m\tau(c_j^2)$. For this, we first compute
    \begin{equation*}
        \sum_{j=1}^m\tau\left(b_j^2\right) = \tau\left((a_0+a_1)^2\right)+\sum_{j=2}^m\tau\left(a_j^2\right) \geq \tau\left(a_1^2\right)+\sum_{j=2}^m\tau\left(a_j^2\right) \geq 1-\delta,
    \end{equation*}
    where we use that $\tau(a_0^2),\tau(a_0a_1)\geq 0$ in the third step. Next, we find
    \begin{align*}
        \sum_{j=1}^m\tau\left(c_j^2\right) &= \sum_{j=1}^m\tau\left(b_j\rhomo b_j\rhomo\right)\\
        &= \sum_{j=1}^m\tau\left(b_j^2\right)-\tau\left(b_j^2(1-\rhomo)\right)-\tau\left(b_j(1-\rhomo)b_j\rhomo\right)\\
        &\geq 1-\delta -4\tau\left(\rho(1-\rhomo)\right)\\
        &\geq 1-\delta-4\eps,
    \end{align*}
    where the last two inequalities are analogous to the steps used in \eqref{eq:c-b-mod-error-est}.

    We can now use \cite[Theorem 1.2]{dlS22orthog} for the POVM $\{c_j\}_{j=1}^m$ to obtain a PVM $\{p_j\}_{j=1}^m$ satisfying
    \begin{equation*}
        \sum_{j=1}^m\norm{c_j-p_j}_2^2\leq 9(\delta+4\eps).
    \end{equation*}
    By the triangle inequality, we find
    \begin{align*}
        \left(\sum_{j=1}^m\norm{a_j-p_j}_2^2\right)^{1/2}&\leq \left(\sum_{j=1}^m\norm{a_j-b_j}_2^2\right)^{1/2}+\left(\sum_{j=1}^m\norm{b_j-c_j}_2^2\right)^{1/2}+\left(\sum_{j=1}^m\norm{c_j-p_j}_2^2\right)^{1/2}\\
        &\leq \eps+2\sqrt{\eps}+3\sqrt{\delta+4\eps}\\
        &\leq 3\sqrt{2\delta + 10\eps},
    \end{align*}
    proving the theorem.
\end{proof}

If we consider the special case where the positive operators in Theorem~\ref{thm:gen-stab} are projections, we get Theorem~\ref{thm:intro_stab} from the introduction, which we restate here for convenience:

\introstab*

\begin{proof}
    Conditions (a) and (b) of Theorem \ref{thm:gen-stab} are clearly satisfied. For condition (c), we observe that
    \begin{equation*}
        \left|1-\tau\left(\sum_{j=1}^mp_j^2\right)\right|=\left|\tau\left(1-\sum_{j=1}^mp_j\right)\right|\leq \norm{1-\sum_{j=1}^mp_j}_2,
    \end{equation*}
    where we use that the $p_j$ are projections in the first step and the Cauchy-Schwarz inequality in the second step. This shows that condition (c) is satisfied with $\delta=\eps$, and the result follows.
\end{proof}

Finally, we present another corollary to Theorem~\ref{thm:gen-stab}, which shows that if the family of positive operators $\{a_j\}_{j=1}^m$ is known to lie below a given PVM, then this additional structure can be preserved when constructing the operators $\{q_j\}_{j=1}^m$. While we do not need this in the remainder of the paper, we believe that it may be of independent interest.

\begin{cor}
    Let $(M,\tau)$ be a tracial von Neumann algebra and fix $0\leq \eps,\delta\leq 1$. Let $\{a_j\}_{j=1}^m$ be a family of positive operators in $M$ and let $\{p_k\}_{k=1}^n$ be a PVM in $M$. Suppose that there is a partition $\Lambda_1,\dots,\Lambda_n$ of $\{1,\dots,m\}$ such that 
    \begin{enumerate}[label=(\alph*)]
        \item for every $1\leq k\leq n$ and $j\in\Lambda_k$: $a_j\leq p_k$,
        \item $\norm{1-\sum_{j=1}^ma_j}_2\leq \eps$,
        \item $\left|1-\tau(\sum_{j=1}^ma_j^2)\right|\leq \delta$.
    \end{enumerate}
    Then there exists a PVM $\{q_j\}_{j=1}^m$ in $M$ such that $\sum_{j=1}^m\norm{a_j-q_j}_2^2\leq\bigO(\eps+\delta)$ and for all $1\leq k\leq n$ and $j\in\Lambda_k$, we have $q_j\leq p_k$. 
\end{cor}
\begin{proof}
    Firstly, we observe that all $a_j$ are elements of the sub-von-Neumann algebra 
    \begin{equation*}
        \tilde{M}=\bigoplus_{k=1}^np_kMp_k,
    \end{equation*}
    and $\tau$ is still a trace on $\tilde{M}$. We can therefore apply Theorem \ref{thm:gen-stab} to the von Neumann algebra $\tilde{M}$ to find a PVM $\{r_j\}_{j=1}^m$ inside $\tilde{M}$ satisfying $\sum_{j=1}^m\norm{a_j-r_j}_2^2\leq\bigO(\eps+\delta)$. 
    
    Next, we denote by $\lambda(j)$ the unique $k$ such that $j\in\Lambda_k$, and we let $\kappa(k)=\min(\Lambda_k)$ and $r_j^{\perp}=r_j(1-p_{\lambda(j)})$. Now for $1\leq j\leq m$, we define
    \begin{equation*}
        q_j = r_jp_{\lambda(j)} + \delta_{j,\kappa(\lambda(j))} \sum_{l=1}^m r_l^{\perp}p_{\lambda(j)}.
    \end{equation*}
    Note that the $p_k$ are central projections in $\tilde M$, and thus $q_j$ is a sum of projections for each $1\leq j\leq m$. One can then compute
    \begin{equation*}
        \sum_{j=1}^m q_j = \sum_{j=1}^m r_jp_{\lambda(j)} + \sum_{k=1}^n \sum_{l=1}^m r_l^{\perp}p_k = \sum_{j=1}^m r_jp_{\lambda(j)} + r_j^{\perp} = \sum_{j=1}^m r_j = 1,
    \end{equation*}
    and since each $q_j$ is a sum of projections, this implies that each $q_j$ is in fact a projection, showing that $\{q_j\}_{j=1}^m$ is a PVM. Moreover, we clearly have that $q_j\leq p_k$ for all $j\in \Lambda_k$. 

    Finally, using the fact that $a_j\leq p_{\lambda(j)}$, we can compute 
      \begin{align*}
          \sum_{j=1}^m\norm{a_j-q_j}_2^2 &= \sum_{j=1}^m\tau\left(\left(a_j-r_jp_{\lambda(j)}-\delta_{j,\kappa(\lambda(j))}\sum_{l=1}^m r_l^{\perp}p_{\lambda(j)}\right)^2\right)\\
          &\leq\sum_{j=1}^m\tau\left((a_j-r_jp_{\lambda(j)})^2\right)+\sum_{k=1}^n\tau\left(\left(\sum_{l=1}^mr_l^{\perp}p_{\lambda(\kappa(k))}\right)^2\right)\\
          &=\sum_{j=1}^m\tau\left((a_j-r_jp_{\lambda(j)})^2\right)+\sum_{k=1}^n\tau\left(\sum_{l=1}^mr_l^{\perp}p_{\lambda(\kappa(k))}\right)\\
          &=\sum_{j=1}^m\tau\left((a_j-r_jp_{\lambda(j)})^2\right)+\sum_{j=1}^m\tau(r_j^{\perp})\\
          &=\sum_{j=1}^m\tau\left((a_j-r_j)^2\right)\\
          &=\sum_{j=1}^m\norm{a_j-r_j}_2^2,
      \end{align*}
      where for the inequality we use that $\tau(ab)\geq 0$ if $a,b\geq 0$. This concludes the proof.
\end{proof}

\section{Approximate regime}\label{sec:analysis_approx}

In this section we will prove our main results, Theorem~\ref{thm:mainreduction} and Corollary~\ref{cor:intro_hardness}. Firstly, using our stability Theorem~\ref{thm:gen-stab}, we will establish the soundness part of Theorem~\ref{thm:mainreduction}, which together with the results from Section~\ref{sec:analysis_perfect} will yield the desired result. Finally, we check that the construction yields a reduction in the sense of Definition~\ref{def:reduction}, which then allows us to deduce Corollary~\ref{cor:intro_hardness}.

We start with the soundness part of Theorem~\ref{thm:mainreduction} for synchronous values:

\begin{lem}\label{lem:sublemapprox}
    There exists a universal constant $\zeta>0$ such that the following holds. Let $\cG=(Q,A,\pi,V)$ be a synchronous game with uniform question distribution, $X(\cG)$ be its corresponding game graph, $t=\abs{Q}$ be the number of questions, and $(X(\cG),t)$ the associated diagonally weighted $t$-independent set game. For sufficiently small $\eps>0$, if $\omega^*_s(\cG)<1-\eps$, then $\omega^*_s((X(\cG),t))<1-\zeta\frac{\eps^2}{t}$.
\end{lem}

\begin{proof}
We prove the lemma in the contrapositive. Assume that the synchronous quantum value of the $t$-independent set game for $X(\cG)$ satisfies $\omega^*_s((X(\cG),t))\geq 1-\zeta \frac{\eps^2}{t}$ for some universal constant $\zeta$ to be determined later in the proof. Following Theorem~\ref{thm:synvNa}, we consider a synchronous strategy $\overline{\cS}$ corresponding to a family of PVMs $\{P_i^{(q,a)}\}_{(q,a)\in Q \times A}$ for $i\in [t]$ in a tracial von Neumann algebra $(M,\tau)$, which satisfies $\omega^*_s((X(\cG),t), \overline{\cS}) \geq 1 - \zeta\frac{\eps^2}{t}$. Note that we are technically not guaranteed that there exists such a strategy, only that there exist strategies $\overline{\cS}_{\eps'}$ with $\omega^*_s((X(\cG),t), \overline{\cS}_{\eps'}) > 1 - \zeta\frac{\eps^2}{t} - \eps'$ for any $\eps' > 0$. However, doing the analysis below for $\overline{\cS}_{\eps'}$ for a fixed $\eps'$ and then letting $\eps' \to 0$ would not change anything, so we omit this notational inconvenience and assume that $\overline{\cS}$ exists.

Our goal is to construct a synchronous strategy $\cS$ for $\cG$ such that $\omega^*(\cG, \cS)\geq 1-\eps$. In fact, the strategy $\cS$ we construct is a synchronous strategy on the same von Neumann algebra $M$. 

Let $\delta = 1 - \omega^*_s((X(\cG),t), \overline{\cS}) \leq \zeta\frac{\eps^2}{t}$ be the losing probability of the strategy $\overline{\cS}$. Consider for a fixed $q$ the set of operators $\{P_i^{(q,a)}\}_{(i,a) \in [t] \times A}$. By Theorem~\ref{thm:intro_stab}, we get for every $q\in Q$ a set of projections $\{Q_i^{(q,a)}\}_{(i,a) \in [t] \times A}$ such that 
$\sum_{i,a} Q_i^{(q,a)}=1$ and
\begin{align}\label{eq:stabtheorem_consequence}
    \sum_{i,a} \norm{P_i^{(q,a)} - Q_i^{(q,a)}}_2^2 \leq \kappa  \norm{\sum_{i,a} P_i^{(q,a)} -1}_2,
\end{align}
for some universal constant $\kappa$. Moreover, the operators $P_i^{(q,a)}$ satisfy
\begin{align}\label{eq:almostorthog_P}
\begin{split}
    \frac{1}{t^2} \sum_{(i,a)\neq (i',a'),q} \tau(P_i^{(q,a)} P_{i'}^{(q,a')}) &= \frac{1}{t^2}\sum_
     {i, a \neq a',q} \tau(P_i^{(q,a)} P_{i}^{(q,a')}) + 
     \frac{1}{t^2}\sum_
     {i\neq i', a,q} \tau(P_i^{(q,a)} P_{i'}^{(q,a)})  \\
     &\qquad\quad + \frac{1}{t^2}\sum_
     {i\neq i', a \neq a',q} \tau(P_i^{(q,a)} P_{i'}^{(q,a')})\\
     &\leq 2\delta,
\end{split}
\end{align}
where the first sum equals zero since $\overline{\cS}$ is a synchronous strategy, and the bound on the other two terms follows from equation~\eqref{eq:loss_indep} by noting that $V(q,q;a,a')=0$ for $a\neq a'$ as the game is synchronous. In order to conclude that the projections $Q_i^{(q,a)}$ are close to the projections $P_i^{(q,a)}$, we now compute the following:
\begin{align*}
    \sum_q \norm{\sum_{i,a} P_i^{(q,a)} -1}_2^2 &= \sum_q \tau\left( \sum_{i,i',a,a'} P_i^{(q,a)}P_{i'}^{(q,a')} + 1 -2\sum_{i,a} P_i^{(q,a)} \right)\\
    &=\sum_q \tau\left( \sum_{(i,a)\neq (i',a')} P_i^{(q,a)}P_{i'}^{(q,a')} + 1 -\sum_{i,a} P_i^{(q,a)} \right) \\
    &\leq 2t^2\delta,
\end{align*}
where we used equation \eqref{eq:almostorthog_P} together with the fact that 
\[
\sum_q \sum_{i,a} P_i^{(q,a)} = \sum_{q,i,a} P_i^{(q,a)} =  \sum_{i}\sum_{q,a} P_i^{(q,a)} = q,
\]
since $\{P_i^{(q,a)}\}_{(q,a)\in Q\times A}$ is a PVM for every $i\in [t]$. Using Jensen's inequality, we thus conclude that
\begin{align}\label{eq:expect_resolution_identity}
    \E_q \norm{\sum_{i,a} P_i^{(q,a)} -1}_2 \leq \sqrt{\E_q \norm{\sum_{i,a} P_i^{(q,a)} -1}_2^2} \leq \sqrt{2t\delta},
\end{align}
where we denote by $\E_q:=\frac{1}{t}\sum_q$ the expectation over $q\in Q$. 

Combining equation \eqref{eq:stabtheorem_consequence} and equation \eqref{eq:expect_resolution_identity}, we thus get
\begin{align}\label{eq:expect_closeness_Q&P}
    \E_q \sum_{i,a} \norm{P_i^{(q,a)} - Q_i^{(q,a)}}_2^2 \leq \kappa  \E_q\norm{\sum_{i,a} P_i^{(q,a)} -1}_2 \leq \kappa \sqrt{2t\delta}.
\end{align}
In other words, the projections $Q_i^{(q,a)}$ are close ``on average over $q\in Q$'' to the projections $P_i^{(q,a)}$. We now construct the operators
\begin{align*}
    Q_q^a \coloneqq \sum_i Q_i^{(q,a)}. 
\end{align*}
Since the projections $Q_i^{(q,a)}$ satisfy $\sum_{i,a} Q_i^{(q,a)} = 1$ by construction, it follows immediately that the set $\{Q_q^a\}_{a\in A}$ is a PVM for every $q\in Q$. 

By Theorem~\ref{thm:synvNa}, we thus get a corresponding synchronous strategy $\cS$ for the synchronous game $\cG$ whose correlation satisfies
\[
p(a,a'\lvert q,q') = \tau(Q_q^a Q_{q'}^{a'}),
\]
for $q,q'\in Q$ and $a,a'\in A$. In particular, the losing probability for the synchronous game $\cG$ incurred by using the synchronous strategy $\cS$ is
\begin{align}\label{eq:loss_approx_synch}
    1-\omega^*(G,\cS)= \E_{q,q'} \sum_{\substack{a,a':\\ V(q,q';a,a')=0}} \tau(Q_q^a Q_{q'}^{a'}).
\end{align}
Note that by construction $\tau(Q_q^a Q_{q'}^{a'})=\sum_{i,i'} \tau\left( Q_i^{(q,a)} Q_{i'}^{(q',a')}\right)$. For the individual terms in this sum, we have
\begin{align*}
    \tau\left( Q_i^{(q,a)} Q_{i'}^{(q',a')}\right)
    &=  \norm{ Q_i^{(q,a)} Q_{i'}^{(q',a')}}_2^2\\
    &= \norm{Q_i^{(q,a)} \left(Q_{i'}^{(q',a')}-P_{i'}^{(q',a')}\right) +  \left(Q_i^{(q,a)} - P_i^{(q,a)}\right)P_{i'}^{(q',a')}+P_i^{(q,a)}P_{i'}^{(q',a')}}_2^2\\
    &\leq  4 \norm{Q_i^{(q,a)} \left(Q_{i'}^{(q',a')}-P_{i'}^{(q',a')}\right)}_2^2 \\
    &\qquad\quad  +  4\norm{\left(Q_i^{(q,a)} - P_i^{(q,a)}\right)P_{i'}^{(q',a')}}_2^2 + 4\norm{P_i^{(q,a)}P_{i'}^{(q',a')}}_2^2,
\end{align*}
where we used that by the triangle inequality $\norm{x+y}_2^2 \leq 2\norm{x}_2^2 + 2 \norm{y}_2^2$ for any operators $x,y$.
Substituting this into equation~\eqref{eq:loss_approx_synch}, we get the following bound on the losing probability of the game $\cG$ when using the strategy $\cS$:
\begin{align*}
    1-\omega^*(G,\cS) &= \E_{q,q'} \sum_{\substack{a,a':\\ V(q,q';a,a')=0}} \tau(Q_q^a Q_{q'}^{a'})\\
    &\leq \frac{4}{t^2} \sum_{\substack{i,i',q,q',a,a':\\ V(q,q';a,a')=0}}  \Big( \norm{Q_i^{(q,a)} \left(Q_{i'}^{(q',a')}-P_{i'}^{(q',a')}\right)}_2^2 \\
    &\qquad\qquad\qquad\qquad\quad +  \norm{\left(Q_i^{(q,a)} - P_i^{(q,a)}\right)P_{i'}^{(q',a')}}_2^2 +  \norm{P_i^{(q,a)}P_{i'}^{(q',a')}}_2^2 \Big).
\end{align*}
We now consider each term separately. For the first term, we have
\begin{align*}
    \frac{4}{t^2} \sum_{\substack{i,i',q,q',a,a':\\ V(q,q';a,a')=0}}   \norm{Q_i^{(q,a)} \left(Q_{i'}^{(q',a')}-P_{i'}^{(q',a')}\right)}_2^2 &\leq
    \frac{4}{t^2} \sum_{\substack{i,i',q,q',a,a'}}  \tau\left(Q_i^{(q,a)} \left(Q_{i'}^{(q',a')}-P_{i'}^{(q',a')}\right)^2\right)\\ 
    &= \frac{4}{t} \sum_{\substack{i',q',a'}}  \tau\left( \left(Q_{i'}^{(q',a')}-P_{i'}^{(q',a')}\right)^2\right)\\ 
    &= \frac{4}{t} \sum_{\substack{i',q',a'}}  \norm{Q_{i'}^{(q',a')}-P_{i'}^{(q',a')}}_2^2 \\
    &\leq 4\kappa \sqrt{2t\delta},
\end{align*}
where the first inequality holds since we just sum over more positive elements, the second equality holds since $\sum_{i,q,a} Q_i^{(q,a)} = t$, and the last inequality follows by equation \eqref{eq:expect_closeness_Q&P}. Completely analogously, using the fact that $\sum_{i',q',a'}P_{i'}^{(q',a')}=t$, we also get for the second term
\begin{align*}
    \frac{4}{t^2} \sum_{\substack{i,i',q,q',a,a':\\ V(q,q';a,a')=0}} \norm{\left(Q_i^{(q,a)} - P_i^{(q,a)}\right)P_{i'}^{(q',a')}}_2^2 
    \leq 4\kappa \sqrt{2t\delta}.
\end{align*}
Finally, for the third term we get
\begin{align*}
    \frac{4}{t^2} \sum_{\substack{i,i',q,q',a,a':\\ V(q,q';a,a')=0}}\tau\left(P_i^{(q,a)}P_{i'}^{(q',a')}\right) &=  \frac{4}{t^2} \sum_{\substack{i\neq i',q,q',a,a':\\ V(q,q';a,a')=0}}\tau\left(P_i^{(q,a)}P_{i'}^{(q',a')}\right) \\
    &\qquad\qquad\qquad\qquad\quad + \frac{4}{t^2} \sum_{\substack{i,q,q',a,a':\\ V(q,q';a,a')=0}}\tau\left(P_i^{(q,a)}P_{i}^{(q',a')}\right) \\
    &\leq 8\delta + \frac{4}{t^2} \sum_{i,(q,a)\neq (q',a')}\tau\left(P_i^{(q,a)}P_{i}^{(q',a')}\right) \\
    &= 8\delta
\end{align*}
where the inequality follows from applying equation~\eqref{eq:loss_indep} to the first sum, and observing that $V(q,q;a,a)=1$ by synchronicity, and thus $(q,a)\neq (q',a')$ in the second sum, and the last equality follows from the fact that $\{P_i^{(q,a)}\}_{(q,a)}$ is a PVM for every $i$.

Putting everything together, we conclude that $1-\omega^*_s(G,\cS) \leq 8\kappa \sqrt{2t\delta} + 8\delta \leq \kappa' \sqrt{t}\sqrt{\delta}$, for some universal constant $\kappa'$ and thus by the definition of $\delta$, 
\begin{align*}
     \omega^*(G,\cS) \geq 1 - \kappa' \sqrt{t} \sqrt{\delta} \geq 1 - \kappa' \sqrt{\zeta}\eps.
\end{align*}
Hence defining the universal constant $\zeta$ as $\zeta = \frac{1}{\kappa^{\prime 2}}$, we get that $\omega^*_s(G,\cS) \geq 1 - \eps$. This finishes the proof.
\end{proof}

We can now complete the proof of our main Theorem~\ref{thm:mainreduction}, which we state here again for convenience:

\mainreduction*

\begin{proof}
    Let $X(\cG)$ be the game graph of $\cG$, and define $\overline{\cG}$ to be the diagonally weighted $t$-independent set game for $X(\cG)$. The first bullet point then follows from the ``only if'' direction of Theorem~\ref{thm:MRV15}, see Remark~\ref{rmk:MRV15}. For the second bullet point, assume by contradiction that $\omega^*(\cG) < 1-\eps$, but $\omega^*(\overline{\cG})\geq 1-\xi\frac{\eps^8}{t^4}$, for some universal constant $\xi$ that will be specified later. We observe that $\overline{\cG}$ is $1/2$-diagonally dominant, so by Theorem~\ref{thm:dlSM23}, $\omega_s^*(\overline{\cG})\geq 1-\kappa\sqrt[4]{2\xi}\frac{\eps^2}{t}$ for some universal constant $\kappa$. On the other hand, Lemma~\ref{lem:sublemapprox} tells us that $\omega_s^*(\overline{\cG})<1-\zeta\frac{\eps^2}{t}$ since $\omega_s^*(\cG)\leq \omega^*(\cG)<1-\epsilon$. Choosing the universal constant $\xi=\frac{1}{2}\frac{\zeta^4}{\kappa^4}$, this yields the desired contradiction, finishing the proof.
\end{proof}

\begin{rmk}
    Note that Lemma \ref{lem:sublemapprox} is concerned with synchronous strategies for synchronous games. This means that every strategy considered in Lemma \ref{lem:sublemapprox} will always provide valid answers when both players are asked the same question. Therefore, the question distribution of $\cG$ on the diagonal is not relevant in Lemma \ref{lem:sublemapprox} and therefore also not relevant in Theorem \ref{thm:mainreduction}. This means that Theorem \ref{thm:mainreduction} still applies when $\cG$ is a $C$-diagonally weighted version of a synchronous game with uniform question distribution, for example.
\end{rmk}

Next, we show that the construction from Theorem~\ref{thm:mainreduction} yields a reduction in the sense of Definition~\ref{def:reduction}.

\begin{cor}\label{cor:hardness_indepset}
Let $0<s<1$ be a fixed constant, let $C$ be the corresponding class of synchronous games from Theorem~\ref{thm:NZ23} and let $t$ be the associated constant number of questions in this family. Let $\overline{C}$ denote the class of $t$-independent set games. Then there exists a constant $0 < \overline{s} < 1$ such that the construction from Theorem~\ref{thm:mainreduction} yields a reduction in the sense of Definition~\ref{def:reduction} from the $(1,s)$-$\mathrm{gap}^*$ problem for $C$ to the $(1,\overline{s})$-$\mathrm{gap}^*$ problem for $\overline{C}$.
\end{cor}

\begin{proof}
First, we construct an explicit map that takes instances of $C$ to instances of $\overline{C}$. Following Definition~\ref{def:reduction} of a reduction, we then show the existence of the required Turing machines which implement this map, as well as completeness and soundness of the reduction.
\,\\

\noindent \textbf{The map. } We map a given synchronous game $\cG_n\in C$ to the diagonally weighted $t$-independent set game $(X(\cG_n),t)\in \overline{C}$ on the game graph $X(\cG_n)$ of $\cG_n$. 
\,\\

\noindent \textbf{Efficient Turing machines. }
    Firstly, we note that the question size is constant for both the family $\{\cG_n\}_n$ as well as the family $\{\overline{\cG}_m\}_m$, and that their distributions are fixed. Therefore, we do not need to define Turing machines to sample questions. 
    
    Let $V$ be a Turing machine associated to $C$ for verifying question and answer pairs as in Definition \ref{def:succinctrep}. We now construct an efficient Turing machine $\overline{V}$ with oracle access to $V$, which checks the predicates of the $t$-independent set games for $X(\cG_n)$. 
    
    For $n \in \mathbb{N}$, questions $i,j \in [t]$ and answers $(q,a),(q',a')\in Q \times A$, the Turing machine $\overline{V}$ is defined as follows: it first compares the bits corresponding to $q$ and $q'$ as well as $a$ and $a'$. This is efficient as the number of bits is constant for the questions and grows as polylog$(n)$ for the answers. Then it uses its oracle access to $V$ in order to compute $V_n(q,q'; a,a')$ and $V_n(q',q;a',a)$. The verifier accepts if and only if
    \begin{itemize}
      \item $i=j$, $q=q'$, $a=a'$, or
      \item $i\neq j$, $q\neq q'$ and both $V_n(q,q'; a,a')$ and $V_n(q',q;a',a)$ evaluate to one.
    \end{itemize}
    This is sufficient to cover all possible cases the verifier is supposed to check. Indeed, when $i=j$, the verifier accepts if and only if the answered vertices are the same. When $i\neq j$, the verifier accepts if and only if the answered vertices are distinct and non-adjacent. Checking that $q\neq q'$ implies that the vertices are distinct. Moreover, they are non-adjacent if and only if both $V_n(a,a'\lvert q,q')$, $V_n(a',a\lvert q',q)$ evaluate to one. On the other hand, if $q=q'$, the verifier immediately rejects. Indeed, in this case the vertices are either the same or adjacent, since whenever $V_n(a,a'\lvert q,q)$, $V_n(a',a\lvert q,q)$ evaluate to one it implies that $a=a'$ by synchronicity.
\,\\

\noindent \textbf{Completeness. } From Lemma \ref{thm:MRV15} and Remark \ref{rmk:MRV15}, it follows that for $\cG_n\in C$, if $\omega^*(\cG_n)=1$, then $\omega^*((X(\cG_n),t))=1$. 
\,\\

\noindent \textbf{Soundness. } Let the constant $\eps$ be $\eps=1-s$. From Theorem~\ref{thm:mainreduction}, if $\omega^*(\cG_n)<1-\eps$ then $\omega^*((X(\cG_n),t))<1-\bigO(\frac{\eps^8}{t^{4}})$. Now let $\overline{s}$ equal $\overline{s}=1-\bigO(\frac{\eps^8}{t^{4}})$. Then it is clear that $\overline{s}<1$. Moreover, $\overline{s}>0$ as we can always choose $t$ large enough, by adding dummy bits to the original questions. 
\,\\

\noindent This finishes the proof.
\end{proof}

We now have all the ingredients to deduce our main complexity-theoretic result Corollary~\ref{cor:intro_hardness}, which we state here again for convenience:

\introhardness*

\begin{proof}
    The result follows from combining Corollary~\ref{cor:hardness_indepset}, Theorem~\ref{thm:NZ23}, and Lemma~\ref{lem:REred}.
\end{proof}

We conclude with an important remark regarding the soundness parameter in Corollary \ref{cor:intro_hardness}. 

\begin{rmk}\label{rmk:trivialsoundness}
    For any fixed $t\in \N$, it is clear that there exists an $s$ for which the $(1,s)$-gap$^*$ problem for $t$-independent set games becomes trivial as it is always possible to win a $t$-independent set game with some small probability. The provers could for instance decide on a set of $t$ vertices and answer according to that set of vertices. They would then at least win whenever the verifier asks them for the synchronous questions, i.e.\ when the questions are the same element in $[t]$. In fact, it is easy to see that this is a general phenomenon for any family of games with fixed question set or fixed answer set.
\end{rmk}

\section{Sharpness of the reduction}\label{sec:sharpness}

In this section, we will show that Lemma \ref{lem:sublemapprox} is optimal in its $t$-dependence. We do this by describing a family of synchronous nonlocal games and concretely constructing a strategy for the corresponding diagonally weighted $t$-independent set games. The nonlocal games we consider may be viewed as luck games, since there is no strategy involved: the players win or lose depending on the questions they receive. However, the corresponding $t$-independent set games still allow the players to strategise. 

\begin{defn}
    Let $k,n\in \N$ with $n\geq 1$, $k\geq 2$. The $(k,n)$-luck game $\cG_{k,n}=([kn],\{1\},\pi_{k,n},V_{k,n})$ is a nonlocal game with question set $[kn]$, answer set $\{1\}$, uniform question distribution $\pi_{k,n}$ and predicate $V_{k,n}$ given by 
    \begin{align*}
        V(x,y;1,1) = 1 \iff \begin{cases}
            &x = y, \text{ or}\\
            &x>n \text{ and } y>n.
        \end{cases}
    \end{align*}
\end{defn}
In these games, the players do not get to make any choices, and they have to get lucky to win. The winning probability is given by
\begin{equation}\label{eq:Gkn-win}
    \omega_s^*(\cG_{k,n})=\omega^*(\cG_{k,n})=\frac{(k-1)^2}{k^2}+\frac{1}{k^2}\frac{1}{n}\leq 1-\frac{1}{k}+\frac{1}{k^2},
\end{equation}
so $\omega^*(\cG_{k,n})\rightarrow1$ as $k\rightarrow\infty$. Moreover, note that all these luck games are synchronous.

Next, we show that this provides a family of synchronous games proving sharpness of the bound in Lemma~\ref{lem:sublemapprox}:

\begin{prop}\label{prop:sharpness}
    For sufficiently small fixed $\eps>0$, there exists a family of synchronous games $\cG_n=(Q_n,A_n,\pi_n,V_n)$ for $n\in \N$ with uniform question distribution, $t_n=|Q_n| \to \infty$, and such that the following holds for all $n\in \N$:
    \begin{itemize}
        \item $\omega_s^*(\cG_n)\leq 1-\eps$, and
        \item $\omega_s^*((X(\cG_n),t_n))\geq 1-2\frac{\eps}{t_n}$,
    \end{itemize}
    where $(X(\cG_n),t_n)$ is the corresponding diagonally weighted independent set game on the game graph $X(\cG_n)$. 
\end{prop}
\begin{proof}
Let $\eps = \frac{1}{k} - \frac{1}{k^2}$ for some fixed, sufficiently large, $k\geq 2$, and note that the result for intermediate values of $\eps$ follows from the fact that $\frac{1}{k} - \frac{1}{k^2} \to 0$ as $k\to\infty$. Consider the $(k,n)$-luck games $\cG_{k,n}$, and note that by \eqref{eq:Gkn-win}, these games satisfy $\omega^*_s(\cG_{k,n})\leq 1 - \frac{1}{k} + \frac{1}{k^2} = 1 - \eps$. 

We now consider the corresponding diagonally weighted $nk$-independent set games $(X(\cG_{k,n}),kn)$, which have question and answer sets equal to $[kn]$. We define a synchronous strategy $\cS_{k,n}=\{P_i^q\}_{i,q=1}^{kn}$ on the von Neumann algebra $\C$ by
\begin{equation*}
    P_i^q=\begin{cases}
        1 & \text{if } i=q\text{ and } i>n,\\
        1 & \text{if } i+n=q\text{ and } i\leq n,\\
        0 & \text{otherwise. }
    \end{cases}
\end{equation*}
$\cS_{k,n}$ is a deterministic strategy, and following Equation \eqref{eq:loss_indep}, its losing probability is given by
\begin{equation*}
    \frac{1}{2(kn)^2}\sum_{\substack{i \neq j,\\ q}} P_i^{q} P_j^{q}+ \frac{1}{2(kn)^2}\sum_{\substack{i\neq j, \\ V(q,q';1,1)=0\, \mathrm{or} \\ V(q',q;1,1)=0}} P_i^{q} P_j^{q'}=\frac{1}{2(kn)^2}\sum_{\substack{i \neq j,\\ q}} P_i^{q} P_j^{q}=\frac{2n}{2(kn)^2}=\frac{1}{k^2}\frac{1}{n},
\end{equation*}
where the second term equals zero since the strategy enforces that $q > n$ when $P_i^q\neq 0$, while $V(q,q';1,1) = 0$ enforces that $q\leq n$ or $q'\leq n$. Since $k\geq 2$, and $t_n = kn$, this shows that 
\begin{align*}
\omega_s^*((X(\cG_{k,n}),t_n)) &\geq 1-\frac{1}{k^2n}\\
&\geq 1 - \frac{2}{k^2n} + \frac{2}{k^3n}\\
&= 1 - \frac{2}{kn}(\frac{1}{k} - \frac{1}{k^2})\\
&= 1 - 2\frac{\eps}{t_n},
\end{align*}
proving the proposition.
\end{proof}

\begin{rmk}\label{rmk:sharpness}
    Proposition~\ref{prop:sharpness} shows that Lemma~\ref{lem:sublemapprox} is optimal in its $t$-dependence, i.e., we cannot replace $t$ in its conclusion by $t^\alpha$ for some $0\leq \alpha < 1$. Indeed, if this were the case, then since $\eps$ is fixed in Proposition~\ref{prop:sharpness} and $t_n\to \infty$, we would be able to find $n\in \N$ such that $1 - \zeta \frac{\eps^2}{t_n^\alpha} \leq 1 - 2\frac{\eps}{t_n}$, in which case the conclusions of Lemma~\ref{lem:sublemapprox} and Proposition~\ref{prop:sharpness} would contradict each other.
    On the other hand, we point out that we do not know whether Theorem~\ref{thm:mainreduction} has optimal $t$-dependence. Firstly, we use Theorem \ref{thm:dlSM23}, and we do not know whether this theorem is optimal. Additionally, it seems hard to rule out the existence of different constructions of game graphs that could work better for the purpose of Theorem~\ref{thm:mainreduction}. 
\end{rmk}

\bibliographystyle{alpha}

\bibliography{biblio}

\end{document}